\documentclass[useAMS,usenatbib]{mn2e}

\usepackage{amsmath}
\usepackage{amssymb}
\usepackage{graphicx}
\usepackage{txfonts}
\usepackage{natbib}
\graphicspath {{./figures/}}

\newcommand{\dd}{\mathrm{d}}
\newcommand{\e}{\mathrm{e}}

\newcommand{\apj}{ApJ}
\newcommand{\apjs}{ApJS}
\newcommand{\apjl}{ApJL}
\newcommand{\aap}{A\&A}
\newcommand{\mnras}{MNRAS}
\newcommand{\prd}{Phys. Rev. D}
\newcommand{\jcap}{J. Cosmology Astroparticle Phys.}

\title[On a novel approach using massive clusters at high redshifts as cosmological probe]{On a novel approach using massive clusters at high redshifts as cosmological probe}

\author[J.-C. Waizmann, S. Ettori and L. Moscardini]{J.-C. Waizmann$^{1,2}$\thanks{E-mail:
jcwaizmann@inaf.oabo.it}, S. Ettori$^{1,2}$ and L. Moscardini$^{3,1,2}$\\
$^{1}$INAF - Osservatorio Astronomico di Bologna, via Ranzani 1, 40127 Bologna, Italy\\
$^{2}$INFN, Sezione di Bologna, viale Berti Pichat 6/2, 40127 Bologna, Italy \\
$^{3}$Dipartimento di Astronomia, Università di Bologna, via Ranzani 1, 40127 Bologna, Italy}
\begin{document}

\date{Accepted 2011 July 20. Received 2011 May 20}

\pagerange{\pageref{firstpage}--\pageref{lastpage}} \pubyear{2011}

\maketitle

\label{firstpage}

\begin{abstract}
In this work we propose a novel method for testing the validity of the fiducial $\Lambda$CDM cosmology by measuring the cumulative distribution function of the most massive haloes in a sample of subvolumes of identical size tiled on the sky at a fixed redshift. The fact that the most massive clusters probe the high-mass tail of the mass function, where the difference between $\Lambda$CDM and alternative cosmological models is strongest, makes our method particularly interesting as a cosmological probe. 
We utilise general extreme value statistics (GEV) to obtain a cumulative distribution function of the most massive objects in a given volume. We sample this distribution function according to the number of patches covered by the survey area for a range of different "test cosmologies" and for differently accurate mass estimations of the haloes. By fitting this sample with the GEV distribution function, we can study which parameters are the most sensitive with respect to the test cosmologies.
We find that the peak of the probability distribution function of the most massive halo is well suited to test the validity of the fiducial $\Lambda$CDM model, once we are able to establish a sufficiently complete large-area survey with $M_{\rm lim}\simeq10^{14.5}\, M_\odot\,h^{-1}$ ($M_{\rm lim}\simeq10^{14}\, M_\odot\,h^{-1}$) at redshifts above $z=1$ ($z=1.5$). Being of cumulative nature the proposed measure is robust and an accuracy of $20-30\%$ in the cluster masses would be sufficient to test for alternative models. Since one only needs the most massive system in each angular patch, this method would be ideally suited as a first fast consistency check before going into a more complex statistical analysis of the observed halo sample.
\end{abstract}


\section{Introduction}
Recently the study of the most massive galaxy clusters in the observable Universe saw an increased interest \citep{ Mantz2008, Cayon2010, Holz2010, Mantz2010, Baldi2011, Hoyle2011, Mortonson2011}, which was mainly initiated by the discovery of the very massive high-redshift cluster XMMU J2235.3−2557  at $z=1.4$ with $M_{200}=(7.3\pm1.3)\times 10^{14}\;M_\odot$ \citep{Mullis2005, Rosati2009, Jee2009}. Those studies mainly concentrated on the consistency of the presence of extremely massive clusters at intermediate and high redshifts with the $\Lambda$CDM concordance model. Particular attention was also given to the impact of non-Gaussianity on the high mass end of cosmological structures \citep{Jimenez2009, Cayon2010, Enqvist2011, Paranjape2011, Sartoris2010}.\\
Moreover, recently \cite{Davis2011} applied general extreme value statistics (GEV) (see e.g. \cite{Fisher1928, Gumbel1958, Coles2001}) to study the probability distribution of the most massive halo in a given volume and \cite{Colombi2011} applied GEV to the statistics of Gaussian random fields. Apart from this, GEV, being relatively wide spread in the environmental (see, e.g., \cite{Katz1992}, \cite{Katz2002}) and the financial sciences (see, e.g., \cite{Embrechts1994}), has seen very few applications in the framework of astrophysics. \cite{Bhavsar1985}, for instance, studied the statistics of the brightest cluster galaxies and \cite{Coles1988} applied GEV on the temperature maxima in the CMB.

In this work, however, we are not interested in the single most massive halo in the Universe, but in recovering the cumulative distribution function (CDF) of the most massive halo in subvolumes by fitting a CDF obtained from GEV and to study its possible discerning power for testing different cosmological models. We will present an attractive, simple and robust method for model testing that could be applied to future wide-field cluster surveys like \textit{EUCLID} \citep{Laureijs2009} or \textit{eROSITA} \citep{Cappelluti2011} for instance.\\
The plan of this paper is as follows. In Sect.~\ref{sec:GEV} we briefly introduce the application of GEV on massive clusters as discussed by \cite{Davis2011}, followed by an introduction to our idea of measuring the underlying CDF for massive clusters in Sect.~\ref{sec:idea}. In Sect.~\ref{sec:models} we study the parameters of the GEV distribution for several cosmological test models. First, for the case of a small fixed comoving volume in Subsect.~\ref{subsec:smallV} and then we extend our method to subvolumes of arbitrary depth in redshift space in Subsect.~\ref{sec:models_realistic}. In order to study the usability of our method, we utilise in Sect.~\ref{sec:sampling} an inverse sampling technique to create observed samples of the underlying CDF distributions for the different cosmologies and recover them by fitting a GEV cumulative distribution function to them. Section~\ref{sec:noise} discusses the robustness of the method with respect to inaccuracies of the mass estimates. We summarise our findings in the conclusions in Sect.~\ref{sec:conclusions} and briefly review the aspects of GEV as it is necessary for this work including a list of the most useful relations in the Appendices \ref{sec:A} and \ref{sec:B}.
\section{The GEV Statistic}\label{sec:GEV}
This section briefly introduces the relations of the GEV relevant for this work. A more detailed discussion can be found in the Appendix~\ref{sec:A}. Following \cite{Davis2011}, the starting point for the application of the GEV statistic is the cumulative distribution function (CDF)
\begin{equation}
P_{\rm GEV}(m)\equiv {\rm Prob.}(m_{\rm max} \le m)
\equiv \int_0^{m}p_{\rm GEV}(m_{\rm max}) \,\dd m_{\rm max},
\end{equation}
which gives the probability of finding a maximum halo mass $m_{\rm max}$ smaller than $m$. In extreme value theory it has been shown that the CDF takes the following functional form \citep{Fisher1928}
\begin{equation}\label{eq:p_gev}
 P_{\mathrm{GEV}}(u) = \exp{\left\lbrace -\left[1+\gamma \left(\frac{u-\alpha}{\beta}\right)\right]^{-1/\gamma}\right\rbrace},
\end{equation}
where $u\equiv\log_{10}m$ is the random variable and $\alpha$, $\beta$ and $\gamma$ are the shift, scale and shape parameter of the distribution, respectively. As shown in \cite{Davis2011}, the parameters in the Poisson limit (on scales $\geq 100\,\textrm{Mpc}\,h^{-1}$ for the application to galaxy clusters)  are found to be 
\begin{eqnarray}\label{eq:parameters}
\gamma = n(>m_0)V-1, \quad \beta =
\frac{(1+\gamma)^{(1+\gamma)}}{\left.\frac{\dd\,n}{\dd\,m}\right|_{m_0}Vm_0\ln 10}, \nonumber \\
\alpha = \log_{10} m_0 - \frac{\beta}{\gamma}[(1+\gamma)^{-\gamma} -1],
\end{eqnarray}
where $m_0$ is the most likely maximum mass, $n(>m_0)$ is the comoving number density of haloes more massive than $m_0$, $V$ is the comoving volume of interest and $\left.\frac{\dd n}{\dd m}\right|_{m_0}$ is the comoving mass function evaluated at $m_0$. The most likely mass, $m_0$, can be found \citep{Davis2011} by performing a root search on
\begin{equation}\label{eq:m0}
\begin{split}
&A \frac{\bar{\rho}_{\rm m} V}{m_0}\sqrt{\frac{a}{2\pi \nu_0}}
\e ^{-a\nu_0/2} \left[ 1+(a\nu_0)^{-p} \right]  \\
& \quad { } -\frac{a}{2} - \frac{1}{2\nu_0} - \frac{ap
 (a\nu_0)^{-(p+1)}}{1+(a\nu_0)^{-p}} + \frac{\nu_0''}{\nu_0'^2} =0.
\end{split}
\end{equation}
Here $\bar{\rho}_{\rm m}=\Omega_{\rm m0}\,\rho_{\rm crit}$ is the mean matter density today, $\nu_0=\left[\delta_{\rm c}/\sigma(m_0,z)\right]^2$ with primes denoting derivatives with respect to $m$ and $A$, $a$ and $p$ are the parameters of the Sheth-Tormen (ST; \cite{Sheth&Tormen1999}) mass function, which we will use throughout the paper.
\section{Introducing the conceptual idea}\label{sec:idea}
\begin{figure}
\centering
\includegraphics[width=1.0\linewidth]{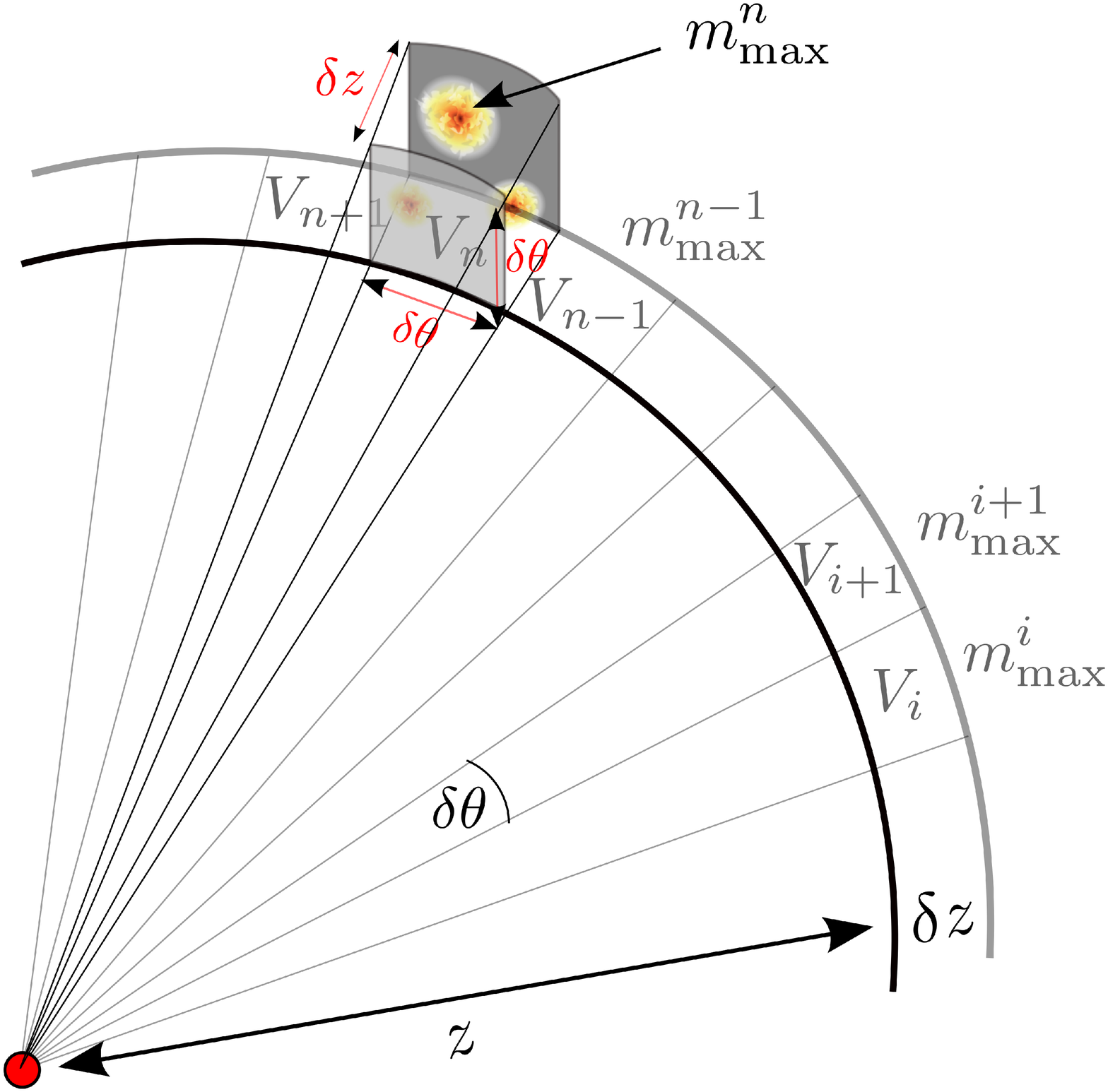}
\caption{Scheme for tiling the sky with subvolumes of size $V_i$ at a redshift $z$ and subsequent measurement of the most massive cluster with mass $m^i_{\rm max}$ in the volume. The faces of the subvolumes at redshift $z$ are assumed to be squares with an angular extent $\delta\theta$ and all subvolumes have an extent in redshift space of $\delta z$.}\label{fig:obs_scheme}
\end{figure}
\begin{figure}
\centering
\includegraphics[width=0.99\linewidth]{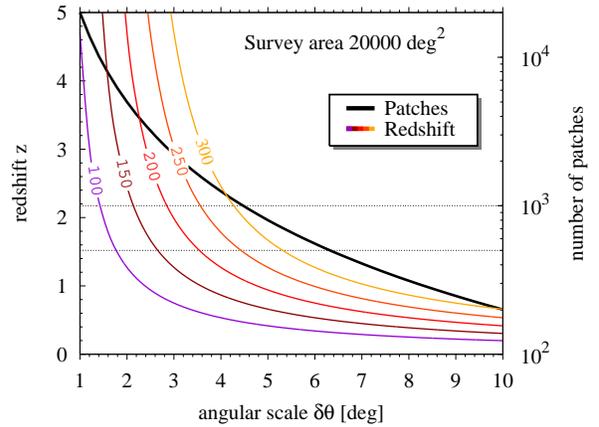}
\caption{Number of square patches of angular scale $\delta\theta$ (right axis of ordinate) as a function of angular scale for a survey area of $20\,000\;\mathrm{deg}^2$ (black line). In addition, the redshift dependence of a fixed comoving transverse distance between $100-300\;\mathrm{Mpc}\, h^{-1}$ (left axis of ordinate) is shown as a function of angular scale for the $\Lambda$CDM cosmology. For better orientation, the grey dotted lines denote the lines of $500$ and $1000$ patches, respectively.}\label{fig:ang_scale}
\end{figure}
In order to observe the CDF, one has to sample a number of subvolumes so that one may compile a sample big enough to recover the underlying distribution function. In the following, we study this approach in the framework of a hypothetical deep wide-field survey covering an area of $20\,000\;\mathrm{deg}^2$ and capable to detect clusters at redshifts above $z=1$.\\
The basic idea, as depicted in the scheme shown in Fig.~\ref{fig:obs_scheme}, is to tile the sky with square patches of angular extent $\delta\theta$ (side length of the patch) and depth $\delta z$, which could be chosen to correspond to a given comoving length for a given cosmological model. In principle, these subvolumes could be placed on a spherical shell anywhere in redshift space, but one has to keep several things in mind.
\begin{enumerate}
\item Naively we want to have as many patches on the sky as possible, to get a better sampling of the underlying distribution.
\item The subvolumes must have a minimum size, such that haloes contained in them can be considered to be uncorrelated and the Poisson limit is valid. Furthermore, larger volumes usually lead on average to a larger $m_\mathrm{max}$, improving the detectability. 
\item  The depth $\delta z$ has to be sufficiently large such that the redshift determination of the most massive cluster is accurate enough to assign the object to the volume.
\item The limiting survey mass should be low enough to allow to completely sample the peak of the distribution as shown in Fig.~\ref{fig:A1}.
\item Moreover, it has to be ensured by the chosen redshift and selection function of the survey that one statistically finds a system in each subvolume.
\end{enumerate}
The number of patches as expected for a hypothetical survey area of $20\,000\;\mathrm{deg}^2$ is shown as a function of angular scale in Fig.~\ref{fig:ang_scale}, showing that in theory one expects $10^2-10^4$ patches. But this number alone is meaningless unless one knows to what redshift a comoving transverse length-scale corresponds for a given angular scale. Therefore, Fig.~\ref{fig:ang_scale} also shows the comoving transverse distance  (see e.g. \cite{Hogg1999})
\begin{equation}
D_{\rm T}=D_{\rm C}\delta\theta,
\end{equation}
where $D_{\rm C}$ is the comoving (line-of-sight) distance, as a function of angular scale for five different length-scales between $100-300\;\mathrm{Mpc}\, h^{-1}$. One directly reads off that is possible to achieve at least a few hundred patches at rather low redshifts and up to several thousands for higher redshift and small length-scales. 
\section{GEV for different cosmological models}\label{sec:models}
\subsection{For a small fixed comoving volume}\label{subsec:smallV}
The attempt of determining the underlying CDF distribution by measuring the mass of the most massive cluster in the subvolumes as outlined in the previous section leads naturally to the question what differences we would expect for such a distribution for different cosmological models.\\
\begin{figure}
\centering
\includegraphics[width=0.9\linewidth]{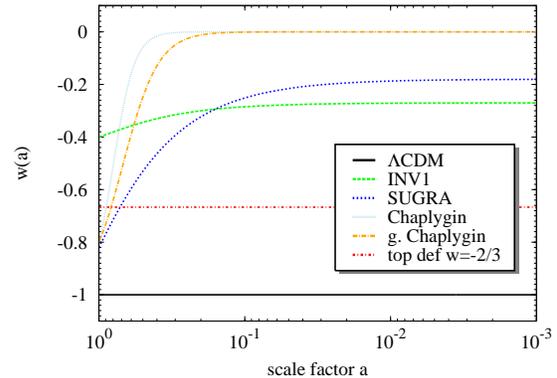}
\caption{Evolution of the equation-of-state parameter $w(a)$ as a function of scale factor $a$.}\label{fig:wz}
\end{figure}
To address this question, we computed the GEV distributions for seven different cosmological models, comprising the fiducial $\Lambda$CDM model with $(h,\Omega_{\Lambda0},\Omega_{\rm m0},\sigma_8)=(0.7,0.73,027,0.81)$ based on the \textit{WMAP} 7-year results \citep{Komatsu2011}, an inverse power law model (INV1), a super-gravity model (SUGRA), the normal and generalised Chaplygin gas, a topological defect model with $w=-2/3$ and an extreme phantom model with $w=-3$. A more detailed description of the used test cosmologies, including the respective linear overdensity thresholds $\delta_{\rm c}$, as well as corresponding references can be found in \cite{Pace2010}. For all models, we use the fiducial $\Lambda$CDM parameters given above with exception of $\sigma_8$ which is scaled according to 
\begin{equation}
\sigma_{8,\mathrm{DE}}=\frac{\delta_{\mathrm{c,DE}}(z=0)}{\delta_{\mathrm{c,\Lambda CDM}}(z=0)}\sigma_8\;.
\end{equation}
\citep{Abramo2007}. The evolution of the equation-of-state parameter $w(a)$ with scale factor $a$ is shown in Fig.~\ref{fig:wz} for all models except the phantom one. It should be noted at this point that the models mentioned above can be referred to as test cases (not necessarily any more consistent with current observations), standing for cosmologies with a substantially different geometric evolution and/or history of structure growth. The fact that our method probes only the high-mass tail of the mass function allows to test for cosmologies that agree more or less on the background level with $\Lambda$CDM, but exhibit substantial differences for the (non-linear) growth history. \newline
\begin{figure}
\centering
\includegraphics[width=\linewidth]{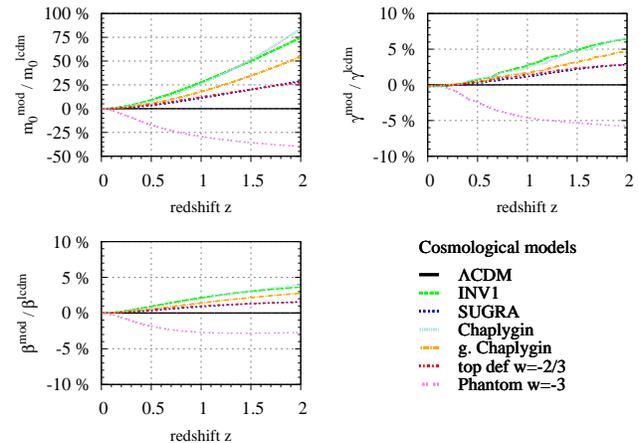}
\caption{Redshift evolution of the ratio with respect to the $\Lambda$CDM model  for $m_0$ (upper-left panel), the GEV shape parameter, $\gamma$, (upper-right panel) and the scale parameter, $\beta$, (lower-left panel) for six different cosmological models, computed for a cubic volume with a comoving side length of $100\,\mathrm{Mpc}\,h^{-1}$.}\label{fig:para_relative}
\end{figure}
As a starting point we compute the three GEV parameters $\alpha$, $\beta$, $\gamma$ and $m_0$ for a fixed cubic comoving volume $V$ with a side length of $L=100\,\mathrm{Mpc}\,h^{-1}$ placed at redshifts in the range of $z\in[0,2]$.  As discussed in \cite{Davis2011}, redshift evolution in $n(>m)$ can be neglected for such a small scale, but it is still big enough to guarantee the validity of the Poisson limit (see also Appendix~\ref{sec:A}). By using the same comoving volume for all models, we neglect for now the different evolution of the cosmic comoving volume $V$ which enters in equation~\eqref{eq:m0} and therefore in all distribution parameters in equation~\eqref{eq:parameters}. In doing so, one lays emphasis on the influence of the different structure growth on the results. \\
In Fig.~\ref{fig:para_relative} we show the results for $\gamma$, $\beta$ and $m_0$; $\alpha$ was left out since it basically coincides with $m_0$. We show in Fig.~\ref{fig:para_relative} the relative deviation from the fiducial $\Lambda$CDM model as a function of redshift for the different cosmological models. Since we normalized all the models to today they all coincide at $z=0$, but start to differ the more we go to higher redshifts. It is directly evident from the upper-left panel that $m_0$ is the most sensitive parameter leading to deviations with respect to $\Lambda$CDM of at least $\sim 10\%$ at $z=1$ and up to $\sim 80\%$ (Chaplygin gas) at $z=2$. The shape and scale parameters $\gamma$ and $\beta$ are much less affected and vary only by a few percent for the given redshift range which will in practise not be measurable. \\
The first result is that one can hope to potentially distinguish different models (with a sufficiently different growth history) at higher redshifts via $m_0$, or equivalently $\alpha$. Of course, it will presumably be harder to control the uncertainties in the mass measurements when going to extremely high redshifts.   
\begin{figure}
\centering
\includegraphics[width=\linewidth]{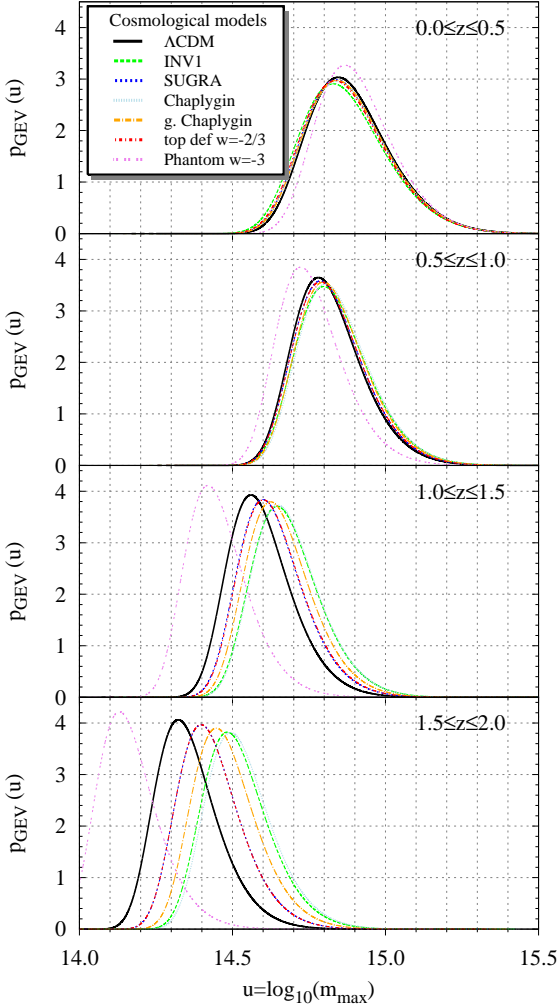}
\caption{PDFs for all seven cosmological models in four different redshift intervals of extent as labelled, assuming an angular scale $\delta\theta=6\,\mathrm{deg}$ of each patch.}\label{fig:multi_vert}
\end{figure}
\subsection{Volumes with significant extent in redshift }\label{sec:models_realistic}
Towards an observationally more realistic case, compared to the small fixed cubic volume with $L=100\,\mathrm{Mpc}\,h^{-1}$, three points have to be addressed first:
\begin{enumerate}
\item In order to assign clusters to the subvolumes, the redshift determination has to be precise enough. In the case of \textit{EUCLID} this will be based on photometric redshifts. The accuracy is estimated to be given by $\sigma_z=0.05(1+z)$ and keeping in mind that $100\,\mathrm{Mpc}\,h^{-1}$ corresponds roughly to $\Delta z=0.03-0.05$ for the redshift range of interest, one must significantly extent the subvolumes in redshift space.
\item An extension in redshift space, however, means that we can no longer neglect the redshift evolution of $n(>m)$ within the volume, since this would lead to a significant overestimation of $m_{\rm max}$ in $V$.
\item Considering that we want to study the effects of different cosmological models on the expected GEV distribution we have to take into account that for a fixed $\Delta z$ and  patch size $\delta\theta$ the volume will be different for each model. Therefore, we expect a contribution by the different evolution of the cosmic volume in the parameters of the distribution as well.
\end{enumerate} 
\begin{figure}
\centering
\includegraphics[width=\linewidth]{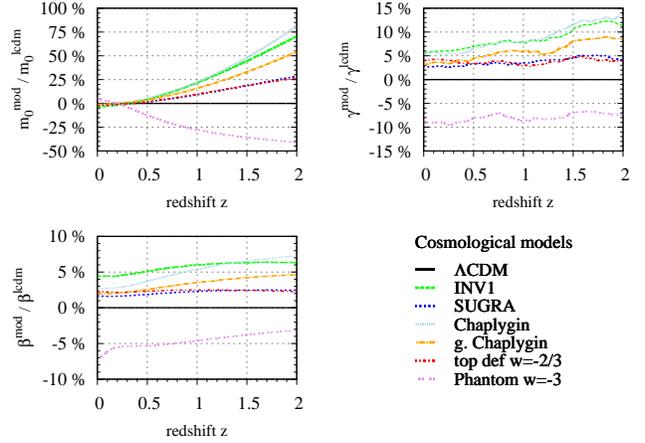}
\caption{Redshift evolution of the ratio with respect to the $\Lambda$CDM model  for $m_0$ (upper-left panel), the GEV shape parameter, $\gamma$, (upper-right panel) and the scale parameter, $\beta$, (lower-left panel) for six different cosmological models, computed for a patch with angular scale $\delta\theta=6\,\mathrm{deg}$ and an extent in redshift space of $\Delta z=0.5$. The redshift $z$ denotes the lower end of the volume.}\label{fig:para_relative_realistic}
\end{figure}
\begin{figure}
\centering
\includegraphics[width=\linewidth]{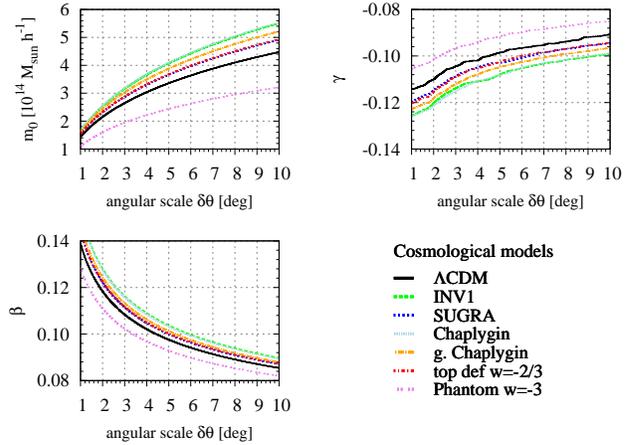}
\caption{Evolution with angular scale $\delta\theta$ of $m_0$ (upper-left panel), the GEV shape parameter, $\gamma$, (upper-right panel) and the scale parameter, $\beta$, (lower-left panel) for seven different cosmological models in the $1.0\leq z\leq 1.5$ interval.}\label{fig:angscale}
\end{figure}
Thus, it seems to be more practical to just fix the patch size $\delta\theta$ and a redshift interval $\Delta z$ instead of fixing the comoving side lengths of $V$.\\
The remaining task is then to define an effective comoving number density $n_{\rm eff}(>m)$ in $\Delta z$. For this, we compute the absolute number $N_V(>m)$ of haloes more massive than $m$ in the volume and define $n_{\rm eff}(>m)=N_V(>m)/V$. Of course it is no longer possible to use the expression from equation~\eqref{eq:m0}, but one has to solve the following equation (see Appendix~\ref{sec:A})
\begin{figure*}
\centering
\includegraphics[width=0.45\linewidth]{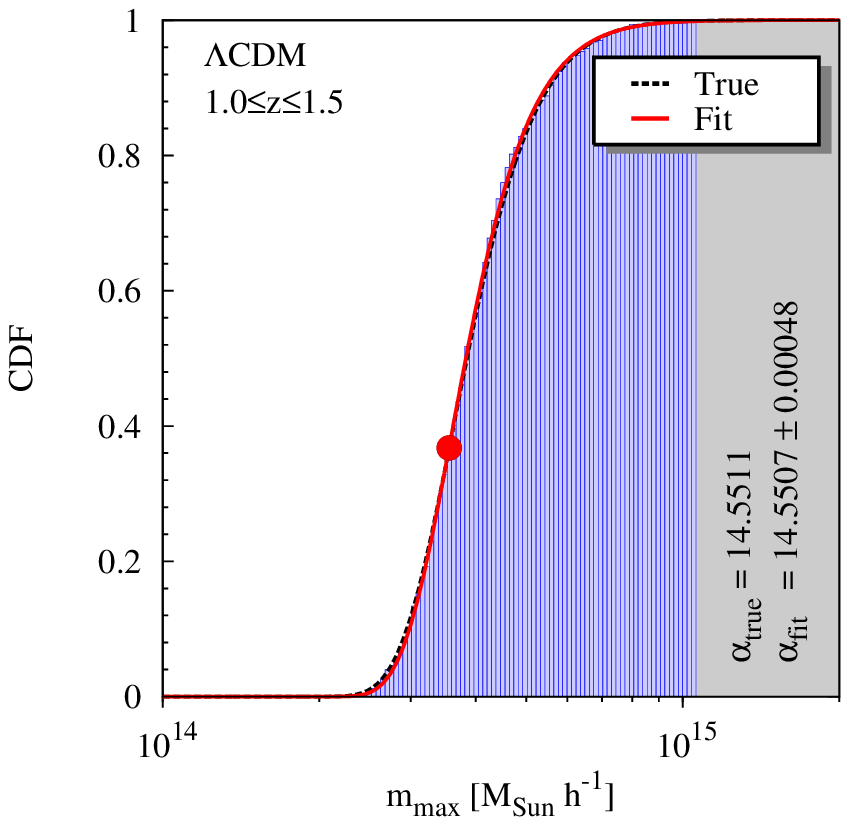}
\includegraphics[width=0.45\linewidth]{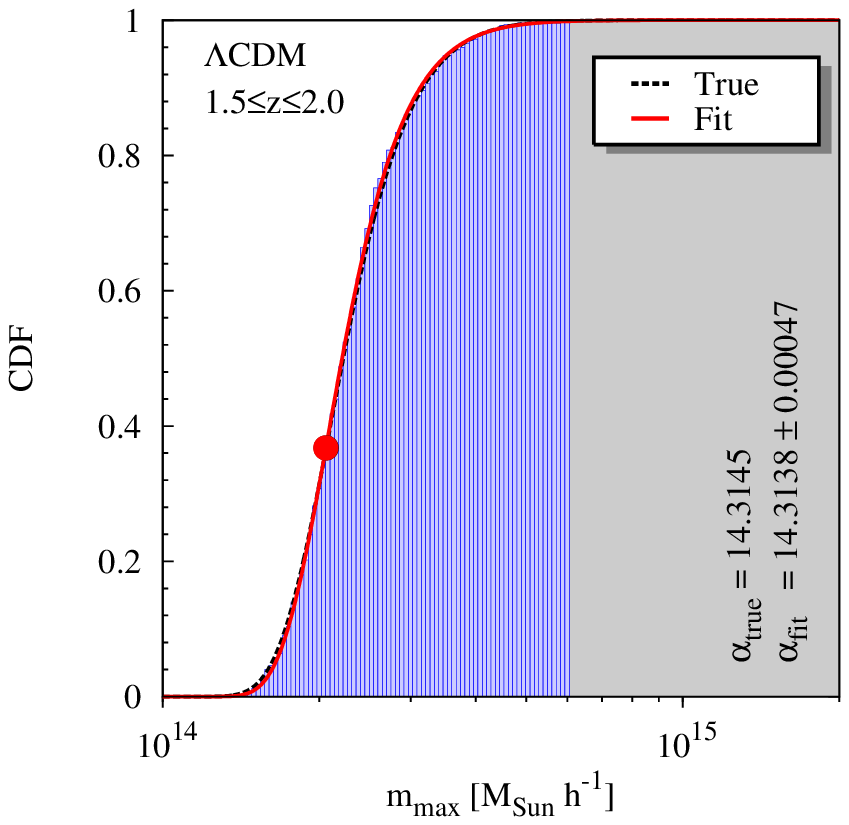}\\
\includegraphics[width=0.45\linewidth]{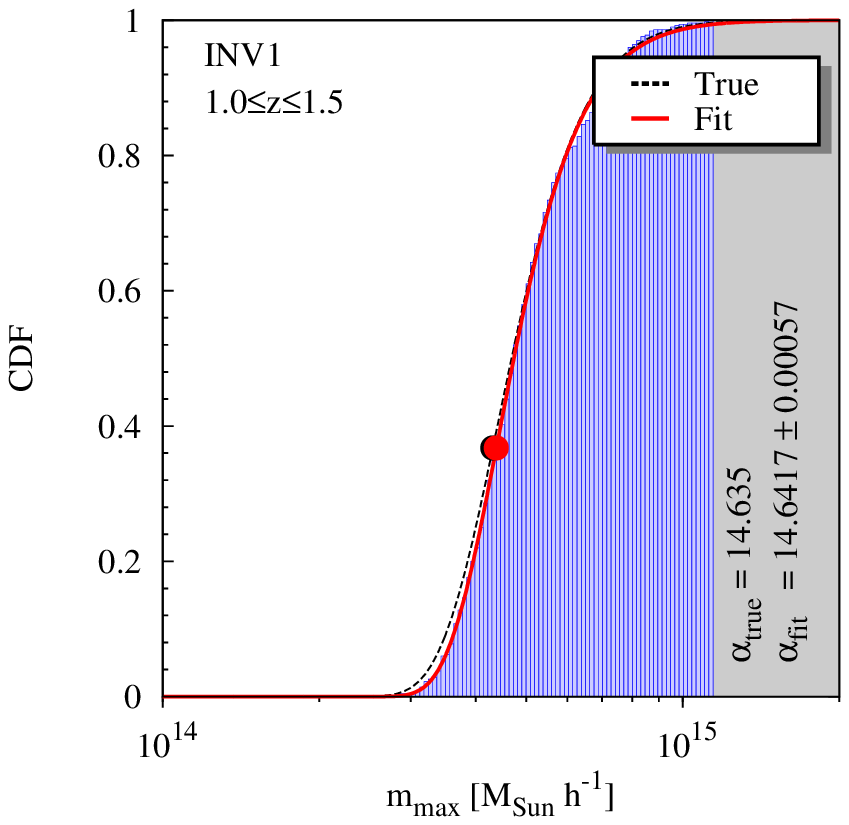}
\includegraphics[width=0.45\linewidth]{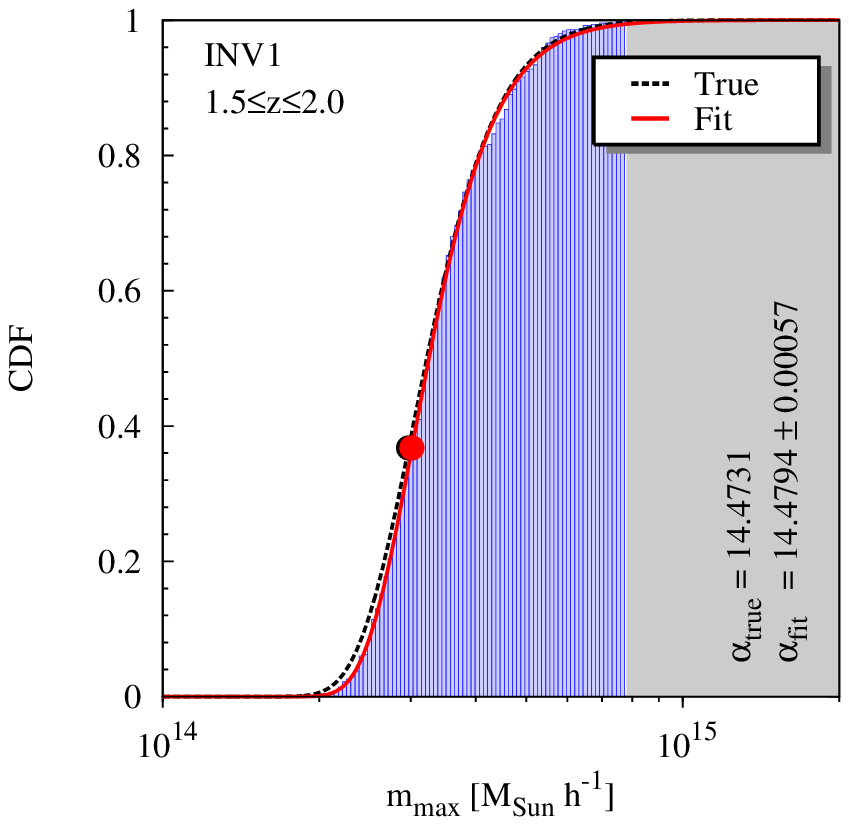}
\caption{Sampled CDF (blue histogram) for the fiducial $\Lambda$CDM model (upper-row) and the INV1 model (lower row), theoretical CDF from the GEV distribution and the fitted CDF (red line) for the redshift interval denoted in the upper-right and an angular patch size of $\delta\theta=6\,\mathrm{deg}$. We observed 500 patches and binned them in 75 sample bins. The numbers in the grey shaded area to the right give the theoretical and fitted values of the GEV $\alpha$-parameter. Only statistical errors are shown.}\label{fig:cdf_lcdm_z1}
\end{figure*}
\begin{equation}\label{eq:m0_num}
 \left.\frac{\dd\,n_{\rm eff}}{\dd\,m}\right|_{m_0}+m_0\left.\frac{\dd ^2\,n_{\rm eff}}{\dd\,m^2}\right|_{m_0}+m_0V\left(\left.\frac{\dd\,n_{\rm eff}}{\dd\,m}\right|_{m_0}\right)^2=0,
\end{equation}
where $\left.\dd\,n_{\rm eff} / \dd\,m \right|_{m_0}$ is the effective mass function evaluated at $m_0$ which is related to the effective number density $n_{\rm eff}(>m)$ via
 \begin{equation}
 \left.\frac{\dd\,n_{\rm eff}}{\dd\,m}\right|_{m_0}=- \left.\frac{\dd\,n_{\rm eff}(>m)}{\dd\,m}\right|_{m_0}.
\end{equation}
Now we can compute the GEV distribution for volumes with a significant extent in redshift space. In Fig.~\ref{fig:multi_vert} the probability distribution functions (PDFs) for all seven previously mentioned cosmological models are shown for four different redshift intervals with $\Delta z =0.5$ placed at $z=\lbrace 0,0.5,1.0,1.5\rbrace$ having an angular patch size of $\delta\theta=6\,\mathrm{deg}$. It can be seen that in the two low-redshift subvolumes the differences between the PDFs are rather small, apart from the very extreme phantom model which has a significantly different volume evolution. At high redshifts, however, the PDFs for all models start to significantly evolve away from each other. This difference is caused by the different growth history and the different evolution of the cosmic volume for each of them. \newline
\begin{figure*}
\centering
\includegraphics[width=0.32\linewidth]{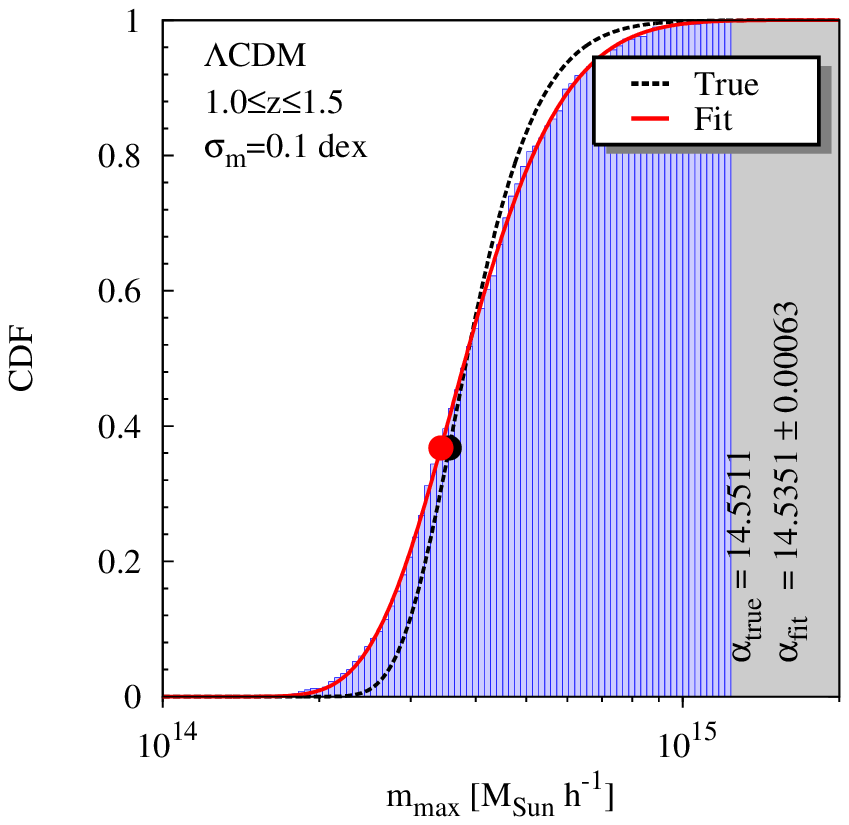}
\includegraphics[width=0.32\linewidth]{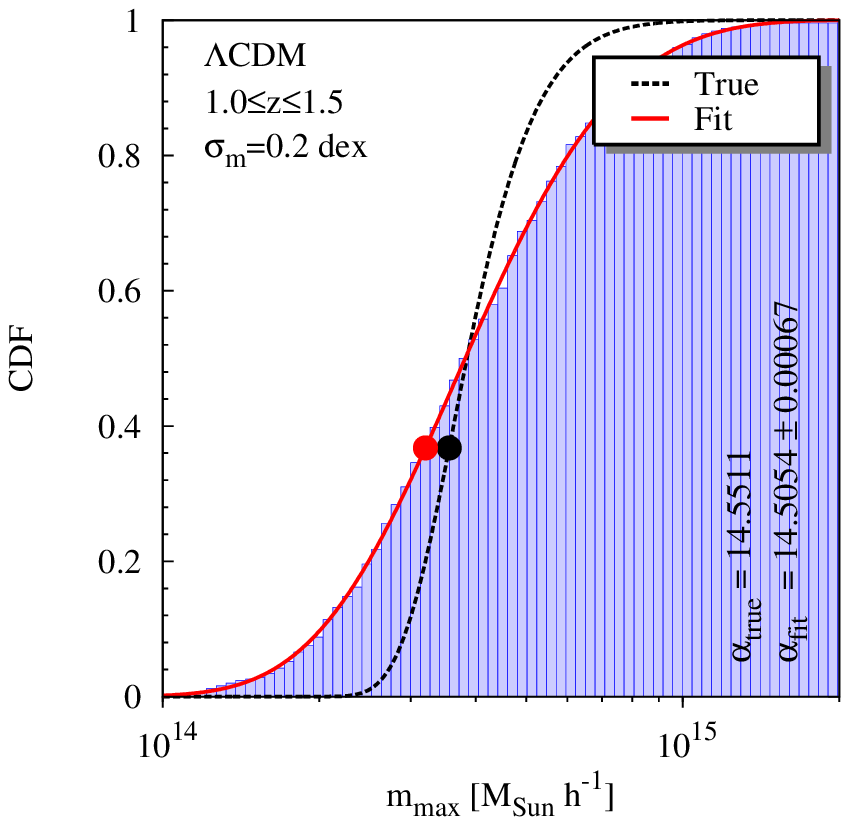}
\includegraphics[width=0.32\linewidth]{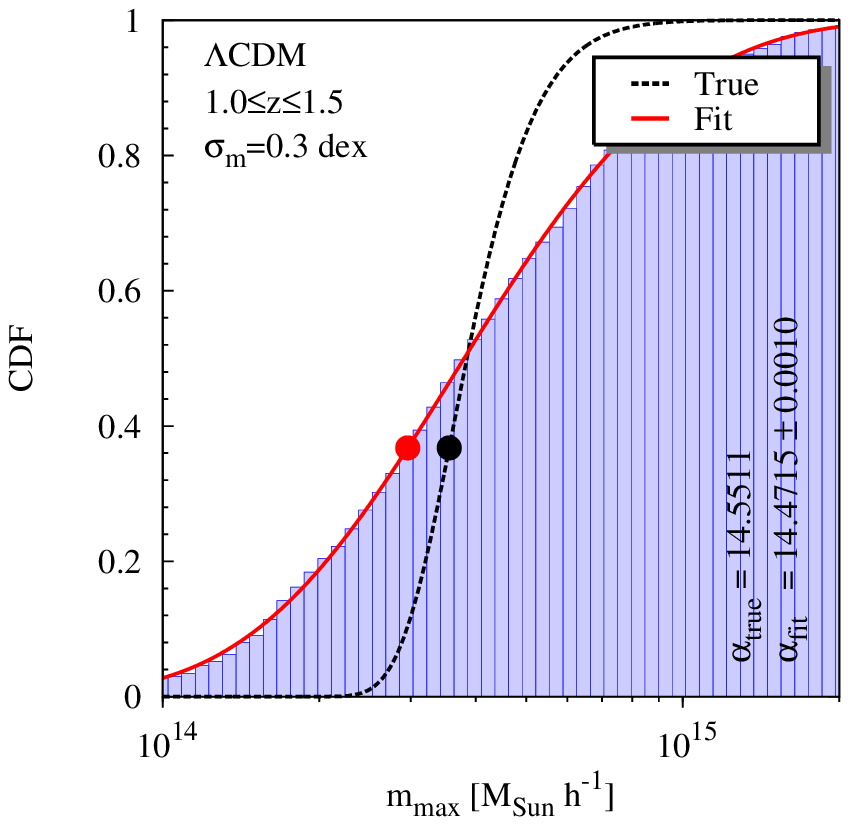}\\
\includegraphics[width=0.32\linewidth]{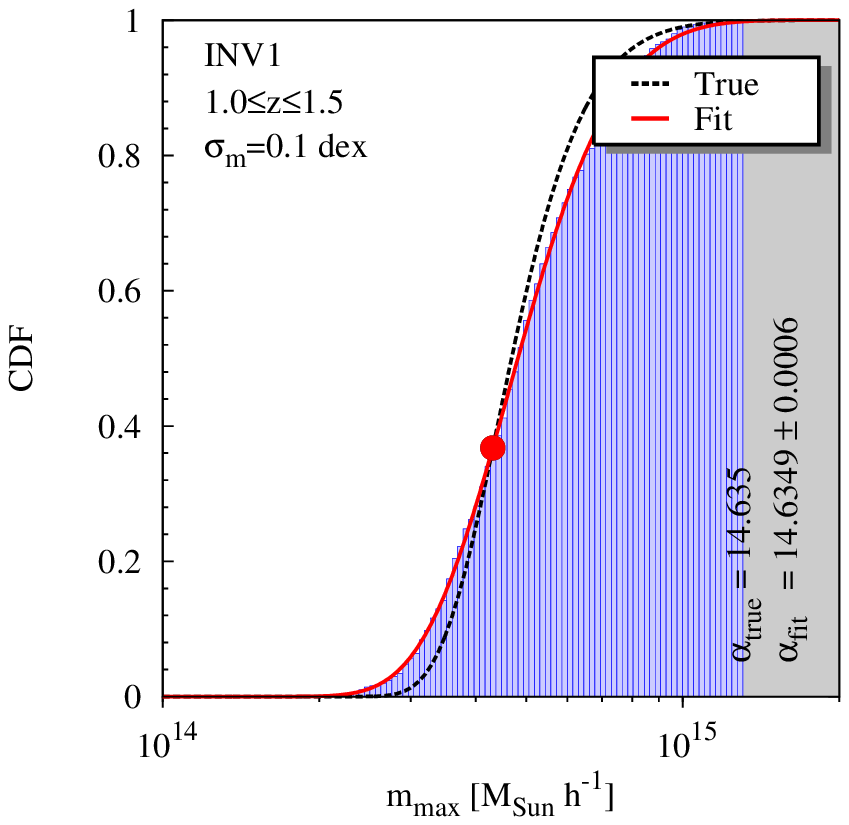}
\includegraphics[width=0.32\linewidth]{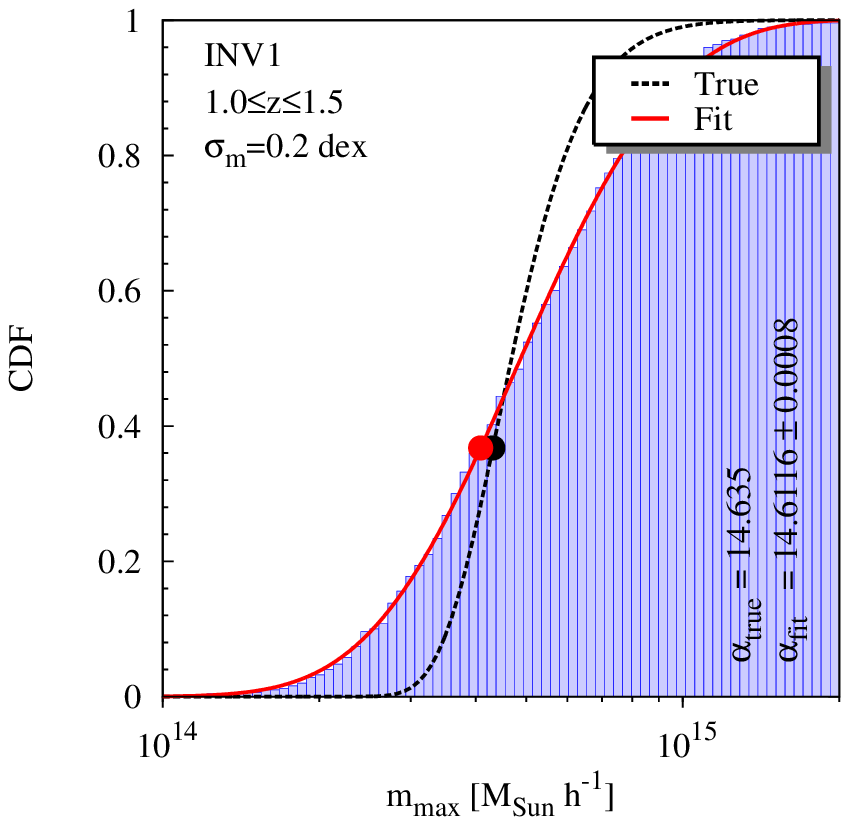}
\includegraphics[width=0.32\linewidth]{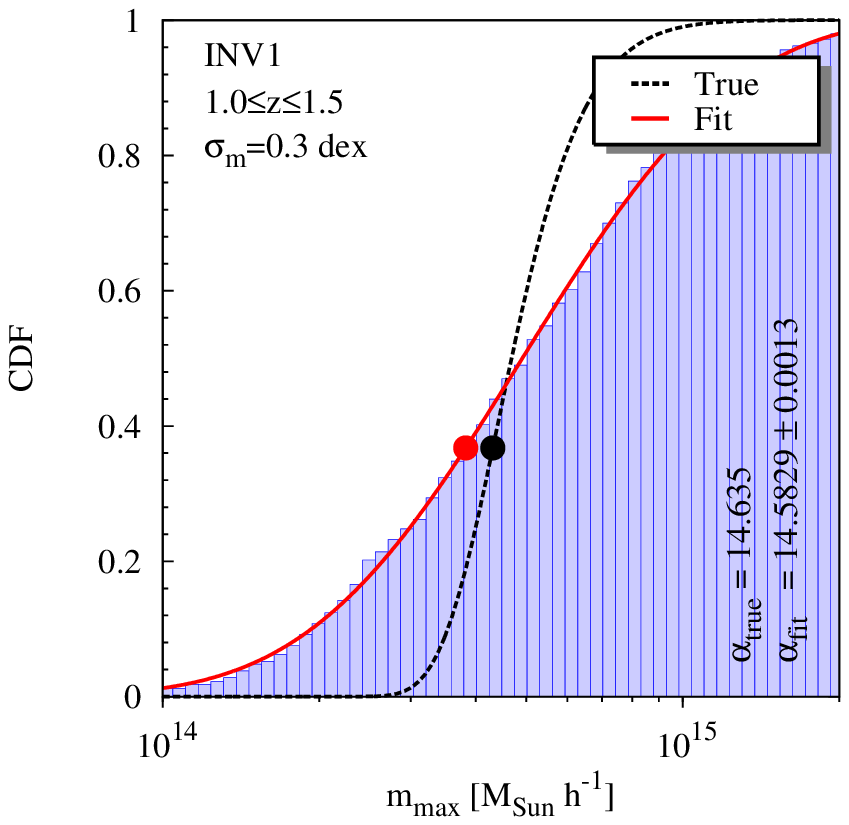}
\caption{Impact of an uncertainty in the mass estimation on the sampled distribution in $1.0\leq z\leq 1.5$: from left to right we assumed a log-normal distributed mass uncertainty with  $\sigma_m= 0.1,\,0.2,\,0.3\;\mathrm{dex}$ for the fiducial $\Lambda$CDM model (upper row) and the INV1 model (lower row). }\label{fig:noisy_cdf_z1}
\end{figure*}
The redshift evolution of the ratio with respect to $\Lambda$CDM of $m_0, \gamma$ and $\beta$ is shown in Fig.~\ref{fig:para_relative_realistic}. The important result is that the peak position of the PDF shows the strongest difference with respect to the fiducial model (more than $10\%$ for $z>1$ and up to $50\%$ for $z>1$), whereas for $\gamma$ and $\beta$ the differences are less than $10\%$ and show only a weak redshift dependence. In Sect.~\ref{sec:noise} we will discuss another reason why $\gamma$ cannot be used to constrain cosmological models. This is a confirmation that trying to measure $m_0$ (or $\alpha$) might be the most promising option.\newline
In Fig.~\ref{fig:angscale} we show the same parameters but as a function of angular patch size $\delta\theta$ in the $1.0\leq z\leq 1.5$ interval, which also corresponds to a change of volume, leading to an increase of $m_0$ with $\delta\theta$, as expected.
\section{Sampling and fitting the GEV distribution}\label{sec:sampling}
In order to address the question of how well one can really observe the cumulative distribution function of the most-massive haloes one has to create a simulated sample of the underlying CDF. An arbitrary CDF can be sampled by means of the inverse sampling technique, requiring an inversion of equation~\eqref{eq:p_gev} for which an analytical form exists. A value $u=\log_{10}\,m$ is drawn by plugging in an equally distributed random number between $0$ and $1$ into the inverted relation. \newline
For the sampling procedure a survey-area of $20\,000\;\mathrm{deg}^2$ was assumed, divided into 500 square patches of $6\times 6\;\mathrm{deg}^2$. We concentrate on two high-redshift bins ($1.0\leq z\leq 1.5$ and $1.5\leq z\leq 2.0$) in which we expect the biggest difference between the fiducial $\Lambda$CDM model and the test cosmologies. After the creation of a sample, which corresponds to the "observation" of the most massive object in each of the $500$ patches, the CDF can be constructed in a straightforward way. First, one sorts the sample with the criteria of increasing mass, secondly one divides the mass interval into a number of bins (75 in this case) and the final remaining task is to count the number of systems with a mass less-equal to the respective mass bin and divide by the total sample-size.  We found the sample and bin numbers to be a good compromise of the stacking in the bins (see Sect.~\ref{sec:noise}), the sampling of the distribution and subvolumes big enough to find clusters in an observable mass range. The samples obtained in this way are shown for $\Lambda$CDM and the INV1 model in Fig.~\ref{fig:cdf_lcdm_z1} by the blue histogram where the black line denotes the theoretical CDF that has been sampled.

Since in a real world application the black line is not known, we have to fit the observed CDF with the theoretical curve from equation~\eqref{eq:p_gev} and determine the parameter values. From the results presented in the previous section it is obvious that only $m_0$, or $\alpha$ have the potential to be used for probing different cosmologies. The shape parameter, $\gamma$, is far too noisy and the scale parameter, $\beta$, only weakly discriminates between the different models. Therefore, we fully concentrate on exploring what can be done by measuring $\alpha$. The values $\alpha_{\mathrm{true}}$ from the theory and the fitted value, $\alpha_{\mathrm{fit}}$, for the $\Lambda$CDM and INV1 models, are given on the right of each panel in Fig.~\ref{fig:cdf_lcdm_z1} and similar plots in the following. The respective values are also depicted as circles on the black line for the theoretical CDF curve and on the red line for the fitted one, respectively. All GEV parameters of the sampled and fitted distributions shown in the following Figs~\ref{fig:cdf_lcdm_z1}-\ref{fig:noisy_cdf_z15_mlim} can be found in Tab.~\ref{tab:parameters}.

The good news is that $500$ patches in $75$mass bins are sufficient to obtain the underlying value of $\alpha$ extremely well for all cases shown in  Fig.~\ref{fig:cdf_lcdm_z1}. The statistical error of the fit is almost negligible and the systematic errors $|\alpha_{\mathrm{true}}-\alpha_{\mathrm{fit}}|$  are small. The difference between the two models shown in Fig.~\ref{fig:cdf_lcdm_z1} is a multiple of the systematic error, such that in this idealised case one would clearly be able to distinguish between the two models. The interesting question now is to study how well the procedure outlined above works as we commit significant errors in measuring the individual masses of the most massive objects, which is discussed in the following section.
\begin{figure*}
\centering
\includegraphics[width=0.32\linewidth]{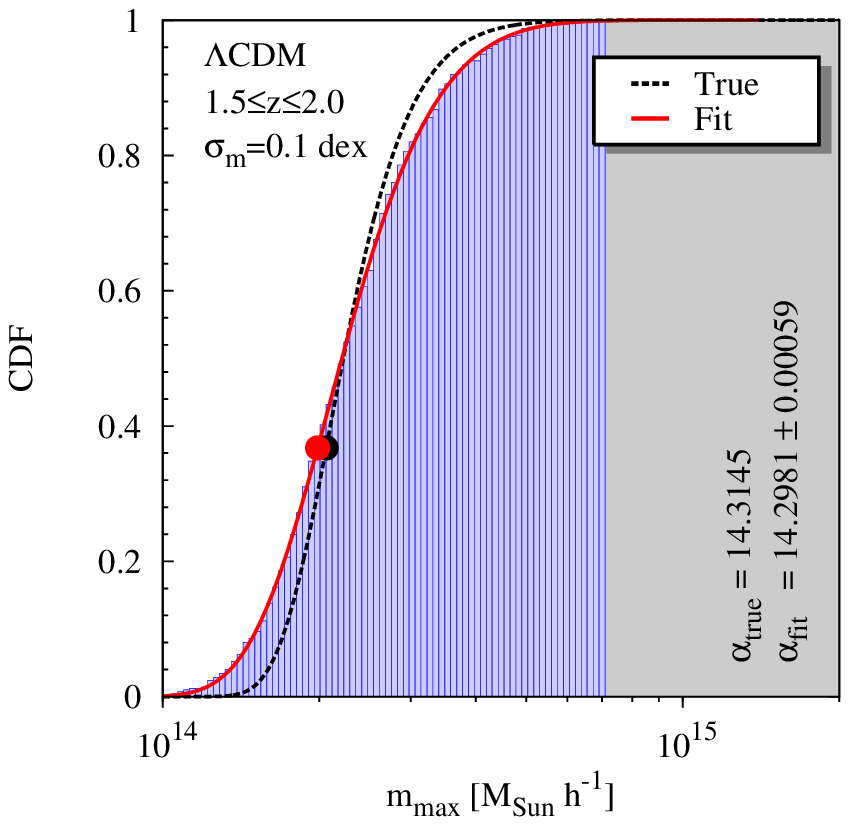}
\includegraphics[width=0.32\linewidth]{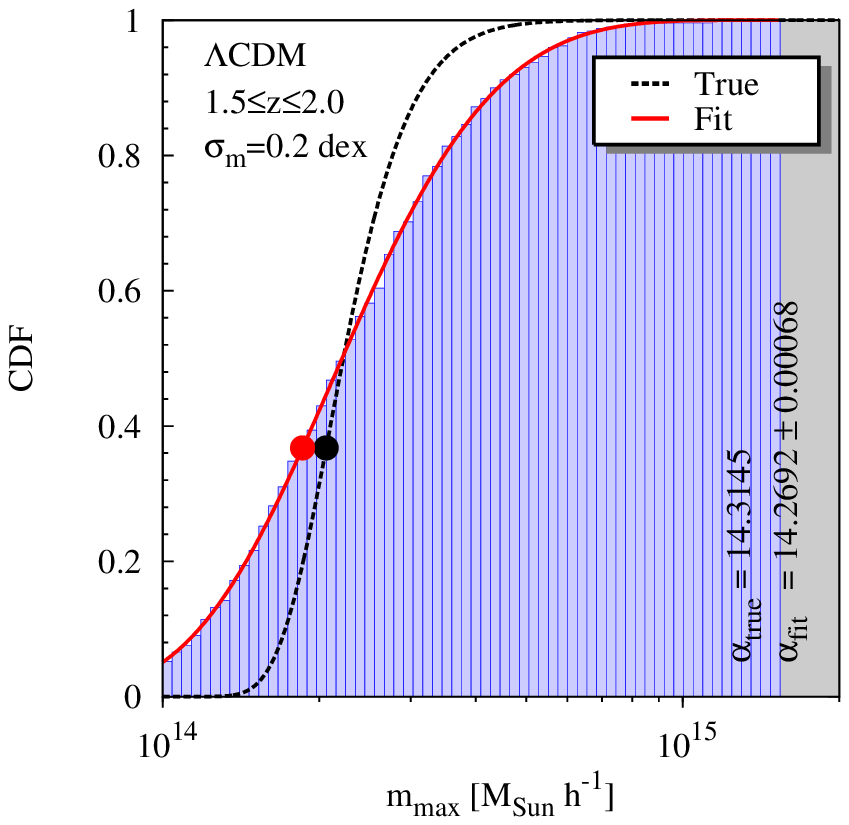}
\includegraphics[width=0.32\linewidth]{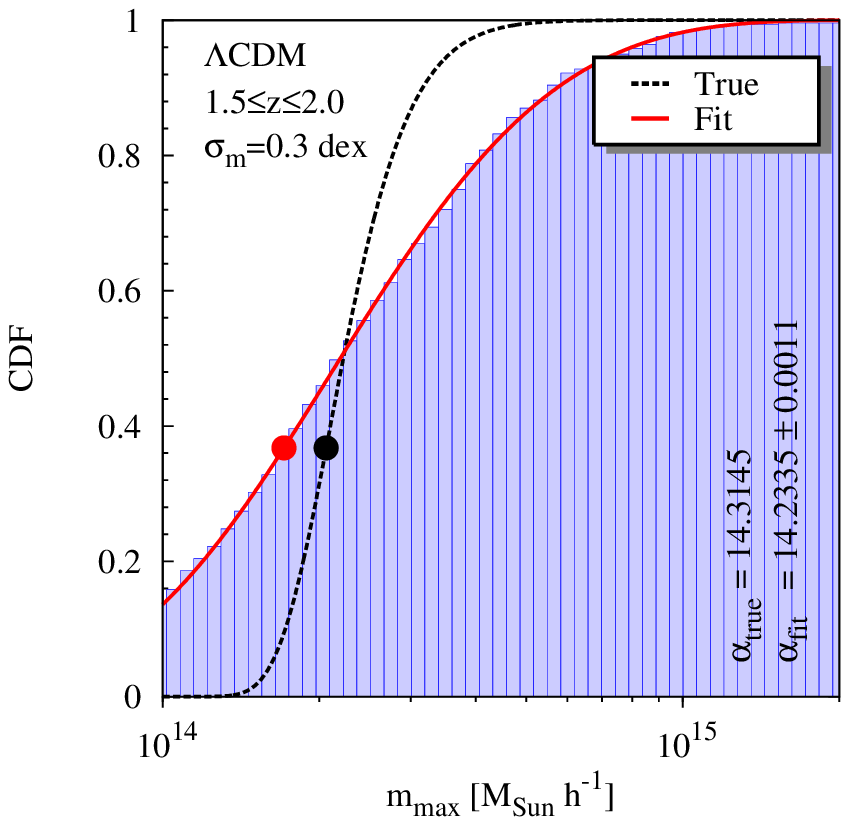}\\
\includegraphics[width=0.32\linewidth]{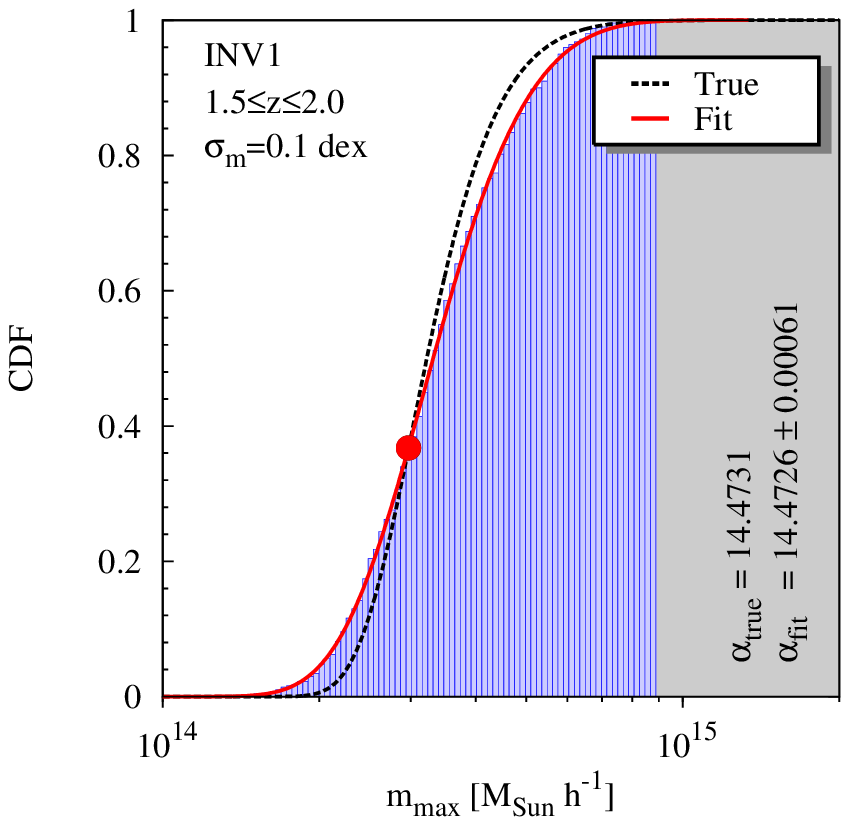}
\includegraphics[width=0.32\linewidth]{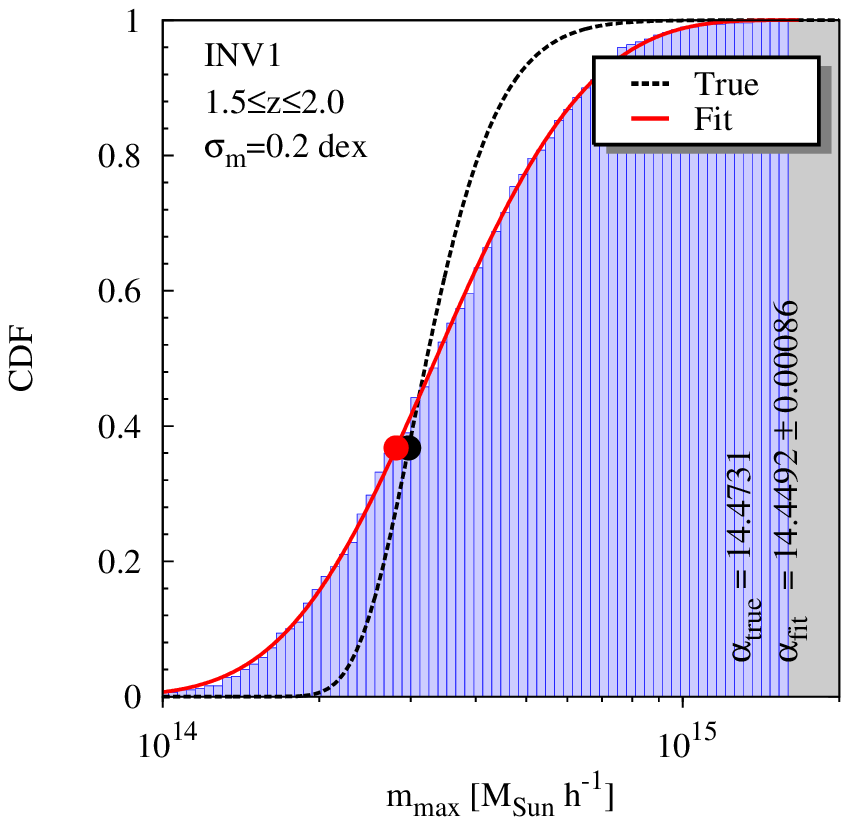}
\includegraphics[width=0.32\linewidth]{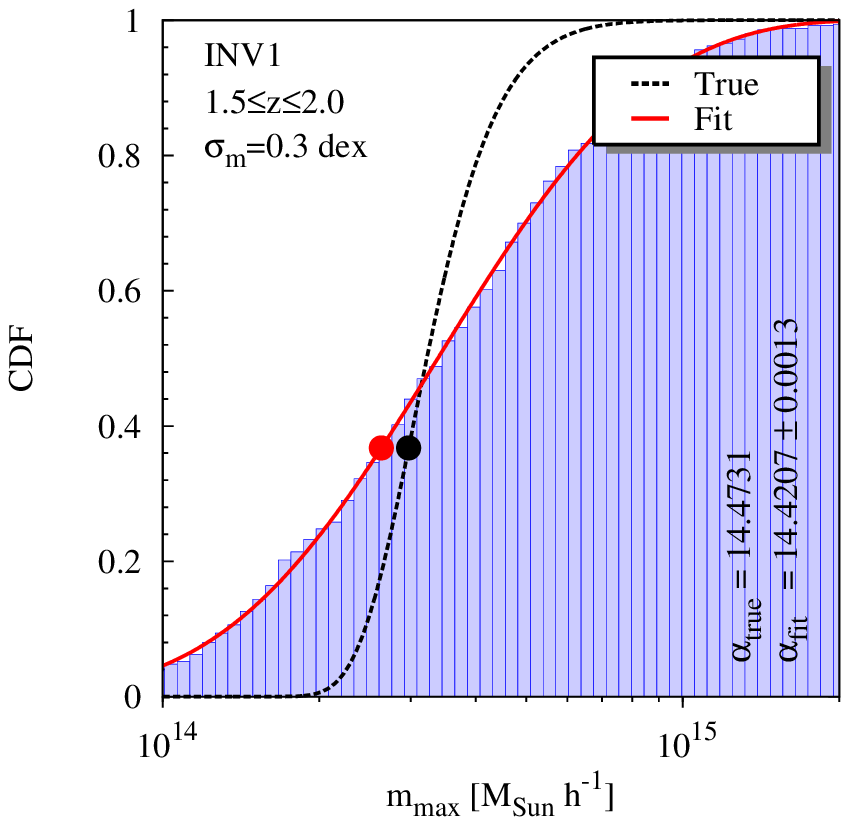}
\caption{Impact of an uncertainty in the mass estimation on the sampled distribution $1.5\leq z\leq 2.0$: from left to right we assumed a log-normal distributed mass uncertainty with  $\sigma_m= 0.1,\,0.2,\,0.3\;\mathrm{dex}$ for the fiducial $\Lambda$CDM model (upper row) and the INV1 model (lower row).}\label{fig:noisy_cdf_z15}
\end{figure*}
\section{Impact of uncertainty in mass estimates}\label{sec:noise}
\begin{figure}
\centering
\includegraphics[width=0.49\linewidth]{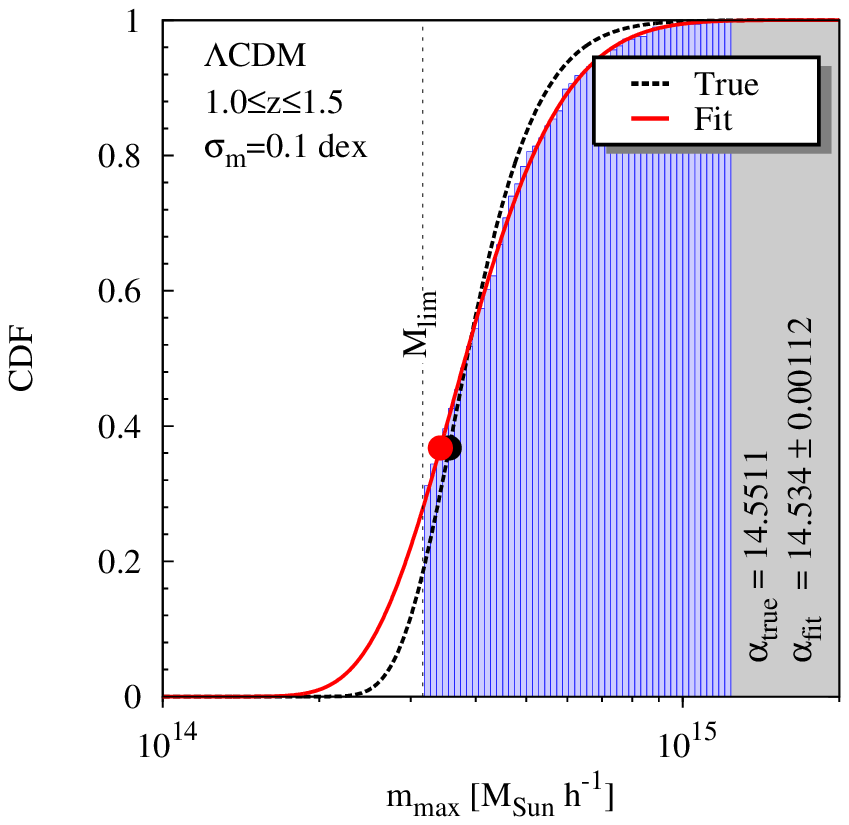}
\includegraphics[width=0.49\linewidth]{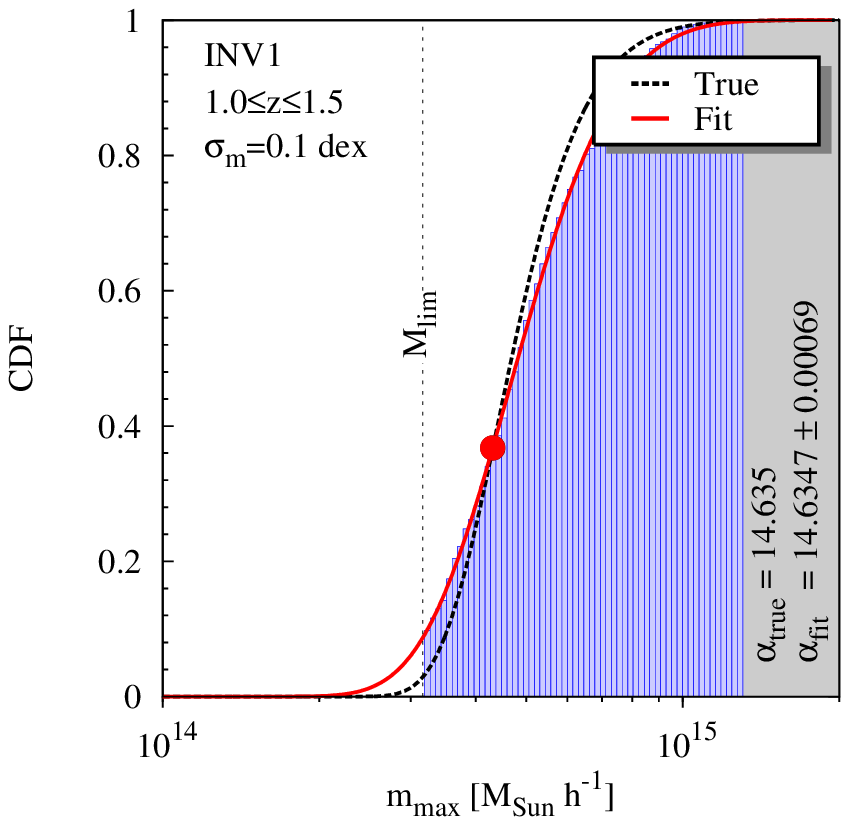}\\
\caption{Impact of limiting survey mass of $M_{\rm lim}\simeq10^{14.5}\, M_\odot\,h^{-1}$ on the recovered CDF for the $1.0\leq z\leq 1.5$ interval for a lognormal distributed mass uncertainty with  $\sigma_m= 0.1\;\mathrm{dex}$ for the fiducial $\Lambda$CDM model (left-hand panel) and the INV1 model (right-hand panel).}\label{fig:noisy_cdf_z15_mlim}
\end{figure}

Unfortunately precise mass determination of galaxy clusters remains challenging, such that in order to understand whether the method proposed in this work is really applicable, it is necessary to study the impact of errors in the mass determination. To get an idea we model the error in the mass determination to be log-normal distributed with a $\sigma_m=\lbrace0.1, 0.2, 0.3\rbrace\,\mathrm{dex}$. When drawing a value $u=\log_{10}\,m$ as discussed in Sect.~\ref{sec:sampling} we change the value by adding a random error obeying the above mentioned log-normal distribution.\newline
The results of this are shown in Fig.~\ref{fig:noisy_cdf_z1} for the redshift interval $1.0\leq z\leq 1.5$ and the $\Lambda$CDM and the INV1 model, where $\sigma_m$ increases from left to right. We decided to show the results for the INV1 model because it is among the models that show the biggest difference with respect to $\Lambda$CDM. As expected an increase in the mass uncertainty substantially alters the shape of the CDF, justifying the previous conclusion that the shape parameter, $\gamma$, cannot be considered as a cosmological probe in this context. However, $\alpha$, which is depicted by the red and black circles for the fitted and true value respectively, seems to be much less affected by the errors in the mass estimates. The fact that the sample is binned automatically implies a stacking of clusters with similar masses, which helps to reduce the scatter. Moreover, it should be mentioned that equation~\eqref{eq:p_gev} delivers good fits in all cases even with substantial mass errors.\\
As a cautionary advice it should be noted that, when following the procedure outlined above, one assumes the observed CDF to be the one of the true most massive clusters; however, observations will always provide the CDF of the most massive observed clusters in the patches. The difference between those two distributions stems from the fact that the most massive observed cluster might be a less massive cluster that scattered up due to errors in the mass determination of the cluster. Incorporating this effect into the above analysis is non-trivial and needs further study before the method could be applied to real data. To some degree the stacking might help to reduce the impact of this effect, but nevertheless the curves of the predicted observed CDF in Figs~\ref{fig:noisy_cdf_z1}-\ref{fig:noisy_cdf_z15_mlim} should systematically shift to higher masses, since even if the mass of the true most massive cluster scatter s down it is inevitable that that a system with a similar mass scatters up. Thus the discerning power of the suggested method can only be used if the impact of mass errors is better understood.\\
A first inspection by eye shows that the difference $\alpha_{\rm fit}$ between the $\Lambda$CDM and the INV1 model is substantial and gets less pronounced the bigger the mass uncertainty is. For the high-redshift sample $1.5\leq z\leq 2.0$ depicted in Fig.~\ref{fig:noisy_cdf_z15} the situation is even better as the difference between the models gets more pronounced.\\
As mentioned above it is crucial to have the limiting mass of the survey in the redshift interval of interest low enough to observe the low-mass tail of the CDF. From Fig.~\ref{fig:noisy_cdf_z15} it can be inferred that a mass of roughly $~10^{14}\,M_\odot\,h^{-1}$  would serve even in the high-redshift case. Due to the robustness of the cumulative measure however, also a more conservative limiting mass of $M_{\rm lim}\simeq10^{14.5}\, M_\odot\,h^{-1}$ would still allow to recover the CDF almost equally well for the $1.0\leq z\leq 1.5$ case as shown in Fig.~\ref{fig:noisy_cdf_z15_mlim}. Such a conservative value of $M_{\rm lim}$ will be in reach of upcoming high-redshift surveys.
\begin{figure*}
\centering
\includegraphics[width=0.32\linewidth]{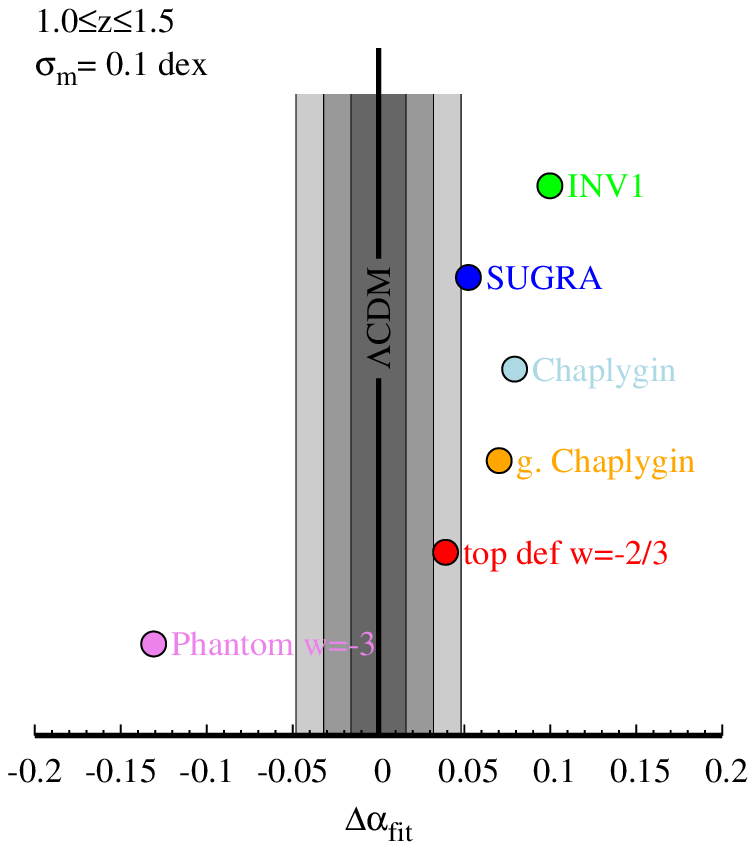}
\includegraphics[width=0.32\linewidth]{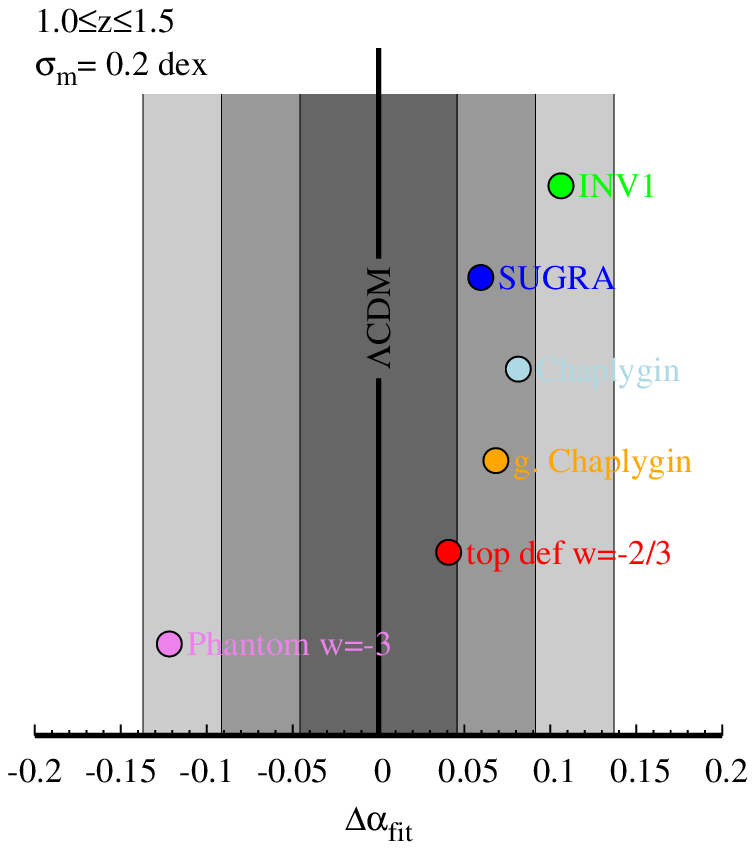}
\includegraphics[width=0.32\linewidth]{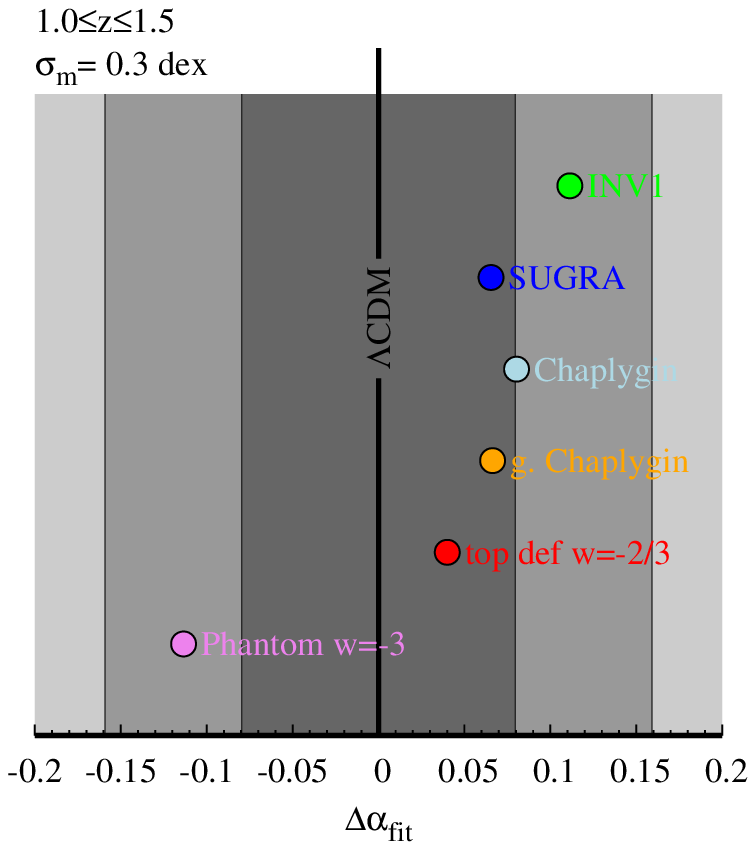}\\
\includegraphics[width=0.32\linewidth]{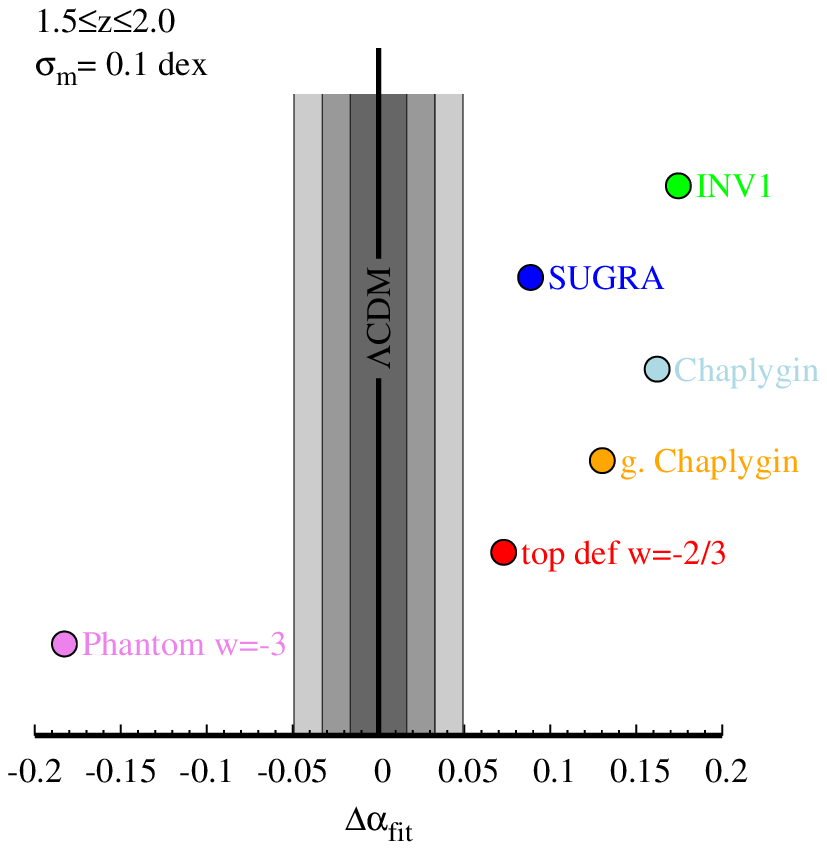}
\includegraphics[width=0.32\linewidth]{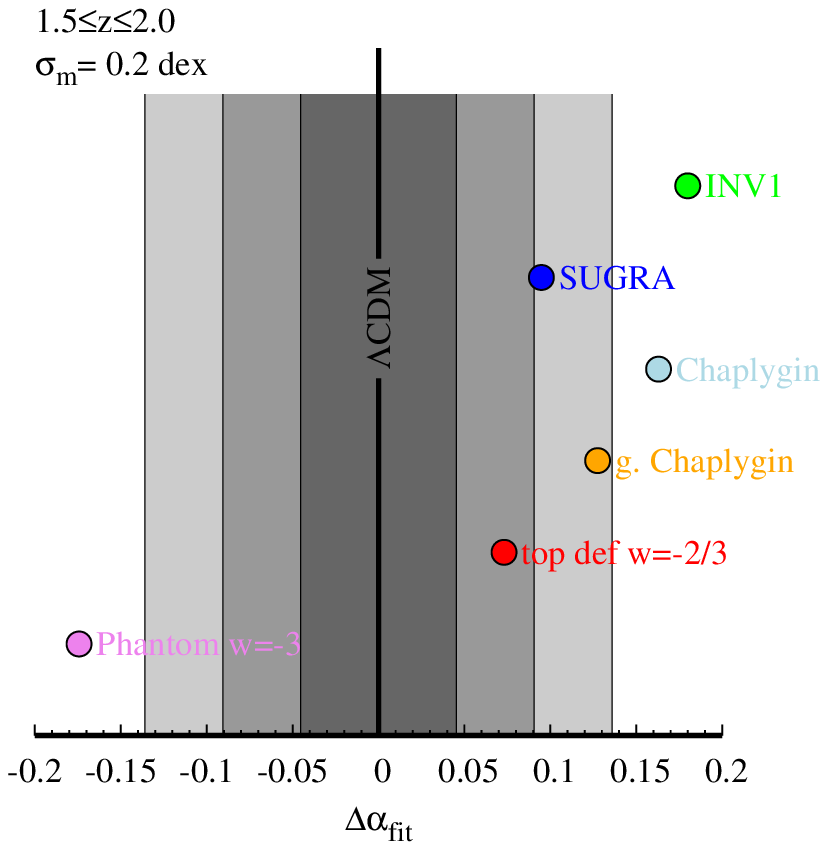}
\includegraphics[width=0.32\linewidth]{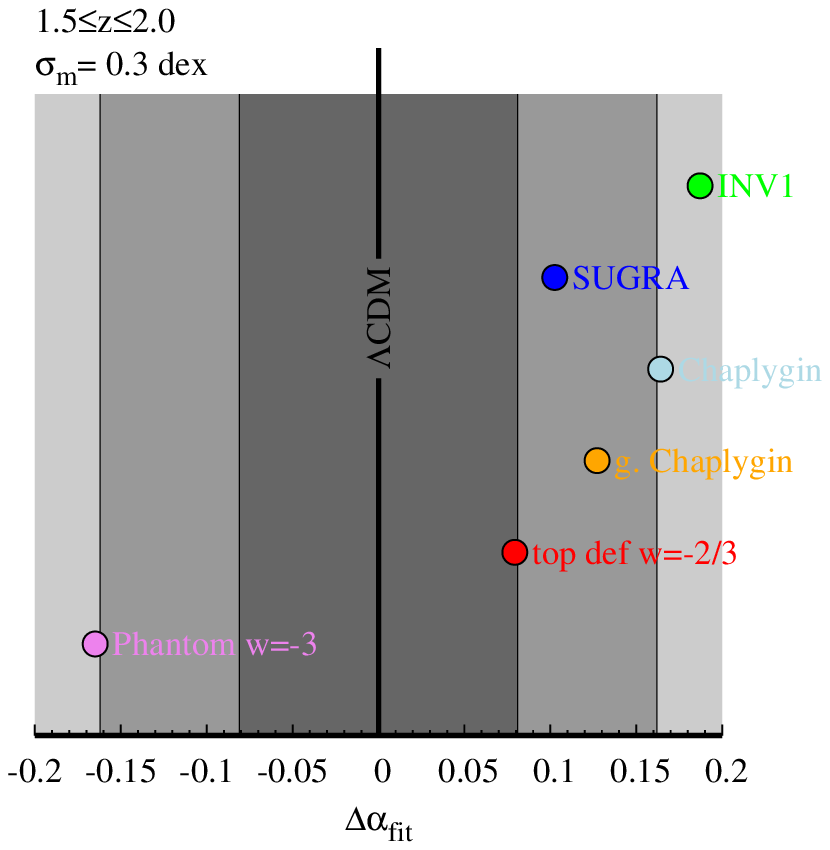}
\caption{Difference between $\alpha_{\rm fit}^{\Lambda \rm CDM}$ and $\alpha_{\rm fit}^{\rm model}$ for all considered models. The grey shaded regions correspond to $1$, $2$ and $3$ times the difference between the theoretical and the fitted value of $\alpha$ for the fiducial cosmology. From left to right the mass uncertainty increases from $\sigma_m=0.1\;\mathrm{dex}$ to $\sigma_m=0.3\;\mathrm{dex}$. The upper row shows the redshift interval of $1.0\leq z\leq 1.5$ and the lower one the interval of $1.5\leq z\leq 2.0$.}\label{fig:diff_z1}
\end{figure*}
The capability to discriminate between different cosmological models of the presented method is presented in Fig.~\ref{fig:diff_z1}, where the differences between the value $\alpha_{\rm fit}$ of the $\Lambda$CDM and the respective test cosmologies are shown. The panels display the differences with an increasing mass uncertainty from left to right, where the upper row presents the results for $1.0\leq z\leq 1.5$ case and the lower row for the $1.5\leq z\leq 2.0$ case. The grey shaded areas show multiples of the systematic error $|\alpha_{\mathrm{true}}-\alpha_{\mathrm{fit}}|$ and we consider a model to be distinguishable if its difference of $\alpha_{\rm fit}$ is more than three times the systematic error. It should be noted that this definition of detectability is extremely conservative, since the systematic error could be accounted for once the impact of the mass errors (lower mass clusters are accidentally mistaken as the most massive systems) is included. Once this is done only the statistical errors would remain which are much smaller and thus would lead to an substantial increase in the discerning power.\\
As expected, the strongest ability to distinguish models is found for the lowest error in the mass estimates of $\sigma_m=0.1\,\mathrm{dex}$ and the high-redshift case as shown in the lower-left panel of Fig.~\ref{fig:diff_z1}. Here, all test models are significantly outside of the grey shaded area, which they enter in the panels for bigger $\sigma_m$ to the right. In the case of $\sigma_m=0.2\,\mathrm{dex}$ the models with the strongest difference with respect to $\Lambda$CDM could still be detected.\\
From an observational point of view the $1.0\leq z\leq 1.5$ case however is of bigger interest since the compilation of the required sample will be easier. Also for this redshift range the $\sigma_m=0.1\,\mathrm{dex}$ case is in principle able to distinguish all test models from $\Lambda$CDM. The more noisy cases could probably still be used as a first $\Lambda$CDM consistency check. Overall these findings show that the method discussed in this work could be of great value for model-testing using a relatively small sample of objects, once the aforementioned treatment of mass errors is understood properly.
\section{Results and summary}\label{sec:conclusions}
In this work we propose a novel method for probing our current cosmological paradigm using the most massive galaxy clusters at high redshifts. The idea is to measure the cumulative distribution function (CDF) of the most massive haloes by tiling the sky with patches of the same area and identical depth in redshift space. We decided for a hypothetical survey of  $20\,000\;\mathrm{deg}^2$, divided into 500 square patches of $6\times 6\;\mathrm{deg}^2$ and $\delta z=0.5$ sampled in $75$ bins, being a good compromise of the stacking in the bins, the sampling of the distribution and subvolumes big enough to find clusters in an observable mass range. Such a survey could e.g. be the proposed \textit{EUCLID} mission or any other large-area survey able to detect galaxy clusters at high-redshifts with a sufficiently low limiting mass. All that is required is the ability to identify the most massive clusters in the individual subvolumes.\\
\begin{table*}
\caption{Compilation of the GEV parameters of the sampled (true) and recovered (fit) distributions as shown in Figs~\ref{fig:cdf_lcdm_z1}-\ref{fig:noisy_cdf_z15_mlim}. The values used in the aforementioned figures are arranged from top to bottom divided by horizontal lines where for the last two rows denoted with ($\star$), corresponding to Fig.~\ref{fig:noisy_cdf_z15_mlim}, a limiting survey mass of $M_{\rm lim}=10^{14.5}\, M_\odot\,h^{-1}$ instead of $M_{\rm lim}=10^{14}\, M_\odot\,h^{-1}$  was assumed. All fits are based on the initial values $\gamma_0=-0.1$, $\beta_0=0.1$ and $\alpha_0=14.5$.}
\begin{tabular}{lcccccccc} \hline
Model & $z$-bin & $\sigma_{\rm m}$ & $\gamma_{\rm true}$ & $\beta_{\rm true}$ & $\alpha_{\rm true}$ & $\gamma_{\rm fit}$ & $\beta_{\rm fit}$ & $\alpha_{\rm fit}$ \\
\hline
$\Lambda$CDM  & $1.0\leq z\leq 1.5$ & & $-0.0957$ & $0.0941$ & $14.5511$ & $-0.0662\pm 0.0093$ & $0.0890\pm 0.0007$ & $14.5507\pm 0.0005$  \\
$\Lambda$CDM  & $1.5\leq z\leq 2.0$ & & $-0.0894$ & $0.0909$ & $14.3145$ & $-0.0564\pm 0.0094$ & $0.0857\pm 0.0007$ & $14.3138\pm 0.0005$   \\
INV1  & $1.0\leq z\leq 1.5$ & & $-0.1051$ & $0.0998$ & $14.635$ & $-0.0499\pm 0.0107$ & $0.0926 \pm 0.0008$ & $14.6417 \pm 0.0006$  \\
INV1  & $1.5\leq z\leq 2.0$ & & $-0.1007$ & $0.0967$ & $14.4731$ & $-0.0449\pm 0.0111$ & $ 0.0895\pm 0.0008$ & $14.4794\pm 0.0006$  \\
\hline
$\Lambda$CDM  & $1.0\leq z\leq 1.5$ & $0.1$ dex & $-0.0957$ & $0.0941$ & $14.5511$ & $-0.1636\pm 0.0085$ & $0.1327\pm 0.0009$ & $14.5351\pm 0.0006$  \\
$\Lambda$CDM  & $1.0\leq z\leq 1.5$ & $0.2$ dex & $-0.0957$ & $0.0941$ & $14.5511$ & $-0.2387\pm 0.0058$ & $0.2174\pm 0.0010$ & $14.5054\pm 0.0007$  \\
$\Lambda$CDM  & $1.0\leq z\leq 1.5$ & $0.3$ dex & $-0.0957$ & $0.0941$ & $14.5511$ & $-0.2637\pm 0.0061$ & $0.3097\pm 0.0015$ & $14.4715\pm 0.0010$  \\
INV1  & $1.0\leq z\leq 1.5$ & $0.1$ dex & $-0.1051$ & $0.0998$ & $14.635$ & $-0.2038\pm 0.0082$ & $0.1362\pm 0.0009$ & $14.6349\pm 0.0006$  \\
INV1  & $1.0\leq z\leq 1.5$ & $0.2$ dex & $-0.1051$ & $0.0998$ & $14.635$ & $-0.2817\pm 0.0068$ & $0.2205\pm 0.0012$ & $14.6116\pm 0.0008$  \\
INV1  & $1.0\leq z\leq 1.5$ & $0.3$ dex & $-0.1051$ & $0.0998$ & $14.635$ & $-0.3024\pm 0.0076$ & $0.3129\pm 0.0018$ & $14.5829\pm 0.0013$  \\
\hline
$\Lambda$CDM  & $1.5\leq z\leq 2.0$ & $0.1$ dex & $-0.0894$ & $0.0909$ & $14.3145$ & $-0.1630\pm 0.0081$ & $0.1308\pm 0.0009$ & $14.2981\pm 0.0006$   \\
$\Lambda$CDM  & $1.5\leq z\leq 2.0$ & $0.2$ dex & $-0.0894$ & $0.0909$ & $14.3145$ & $-0.2392\pm 0.0058$ & $0.2158\pm 0.0010$ & $14.2692\pm 0.0007$   \\
$\Lambda$CDM  & $1.5\leq z\leq 2.0$ & $0.3$ dex & $-0.0894$ & $0.0909$ & $14.3145$ & $-0.2629\pm 0.0064$ & $0.3089\pm 0.0015$ & $14.2335\pm 0.0011$   \\
INV1  & $1.5\leq z\leq 2.0$ & $0.1$ dex & $-0.1007$ & $0.0967$ & $14.4731$ & $-0.2057\pm 0.0083$ & $0.1345\pm 0.0008$ & $14.4726\pm 0.0006$  \\
INV1  & $1.5\leq z\leq 2.0$ & $0.2$ dex & $-0.1007$ & $0.0967$ & $14.4731$ & $-0.2852\pm 0.0074$ & $0.2196\pm 0.0013$ & $14.4492\pm 0.0008$  \\
INV1  & $1.5\leq z\leq 2.0$ & $0.3$ dex & $-0.1007$ & $0.0967$ & $14.4731$ & $-0.3067\pm 0.0078$ & $0.3123\pm 0.0019$ & $14.4207\pm 0.0013$  \\
\hline
$\Lambda$CDM$^\star$  & $1.0\leq z\leq 1.5$ & $0.1$ dex & $-0.0957$ & $0.0941$ & $14.5511$ & $-0.1683\pm 0.0165$ & $0.1342\pm 0.0026$ & $14.534\pm 0.0011$  \\
INV1$^\star$  & $1.0\leq z\leq 1.5$ & $0.1$ dex & $-0.1051$ & $0.0998$ & $14.635$ & $-0.2101\pm 0.0108$ & $0.1371\pm 0.0013$ & $14.6347\pm 0.0007$  \\
\hline
\end{tabular}\label{tab:parameters}
\end{table*}
The method that has been presented in this work has several appealing advantages:
\begin{enumerate}
\item Using a cumulative measure is robust and the binning corresponds to a stacking of clusters of similar mass, reducing the impact of errors in the mass estimation. 
\item We are probing the high-mass tail of the mass function, well above the non-linear mass scale, where the difference between $\Lambda$CDM and alternative cosmological models is strongest.
\item A limiting mass $M_{\rm lim}\simeq10^{14}\, M_\odot\,h^{-1}$ is sufficient to sample the CDF even at high redshifts. For the $1.0\leq z\leq 1.5$ case also $M_{\rm lim}\simeq10^{14.5}\, M_\odot\,h^{-1}$ would be sufficient to recover the CDF, which would definitely be achievable by future wide-area surveys. 
\end{enumerate}
In order to measure the CDF, we fit a function obtained from generalised extreme value theory with three free parameters $\gamma$, $\beta$ and $\alpha$, which are the shape, scale and shift parameters, where the latter is closely related to the position of the peak of the underlying probability density function. Out of these three quantities the shift parameter, $\alpha$, is the most sensitive one for model testing, whereas the shape and scale parameters are not sensitive enough and depend only weakly on redshift. Thus, we propose to use $\alpha$ as discerning measure between cosmological models.\\
We show that by placing the patches at redshift $z=1.0$ assuming an error in the mass estimation of $\sigma_m=0.1\,\mathrm{dex}$ one can distinguish all models considered in this work clearly from the fiducial $\Lambda$CDM case.  Such an accuracy of  $\sigma_m=0.1\,\mathrm{dex}$ in mass could be achieved by combining lensing\footnote{For the $1.5\leq z\leq 2.0$ case mass estimates from lensing can not be utilised.} and X-ray data. Going to higher redshifts, like $z=1.5$ improves the discerning power further. Unfortunately, it is very doubtful that a sufficient mass accuracy can be achieved at such high redshifts by any currently available method. For larger errors in the mass estimation the discerning power is more and more reduced due the stronger impact of the convolution of the theoretical CDF with the distribution of the mass errors. This is also another reason why the shape parameter, $\gamma$ can not be used as discerning measure for different cosmological models. It should also be noted that by choosing the systematic error as measure for the discerning power we decided for a very conservative criterion which could, by correcting for the systematic error, be reduced to the much smaller statistical error.\\

The main focus of this work was to present a novel method for probing the concordance cosmology using massive galaxy clusters. We demonstrated the potential of the method with a collection of toy-models, leading to encouraging results. The simplicity of the suggested scheme makes it in principle a perfect first test of $\Lambda$CDM, before a more demanding reconstruction of e.g. the halo mass function can be achieved. However, before this method can be applied in a real world scenario it is necessary to better understand the real impact of errors in the mass measurements on the observed sample. Particularly mistaking lower mass clusters as the true most massive system might cause a systematic shift to higher masses in the observed CDF undermining the potential as a cosmological probe if not accounted for. Therefore this work can be considered as a first step to an applicable method for model testing. The next step will be to optimise the method further for application on cluster surveys and to better understand the impact of errors in the mass measurements as well as to study the discerning power in more detail for more realistic contenders of $\Lambda$CDM as well as to extend it to non-Gaussian models. 

\section*{Acknowledgments}
We acknowledge financial contributions from contracts ASI-INAF I/023/05/0, ASI-INAF I/088/06/0, ASI I/016/07/0 COFIS, ASI Euclid-DUNE I/064/08/0, ASI-Uni Bologna-Astronomy Dept. Euclid-NIS I/039/10/0, and PRIN MIUR Dark energy and cosmology with large galaxy surveys. Furthermore, the first author would like to thank Matthias Bartelmann for the helpful discussions during a short visit at the \textit{Institut f\"{u}r Theoretische Astrophysik (ITA), Universit\"{a}t Heidelberg.} We also would like to thank the referee for his comments that helped to improve this manuscript.

\bibliographystyle{mn2e}

\appendix

\section{Applying GEV on the most massive haloes}\label{sec:A}
In this appendix we outline the derivation of the most important relations for the application of GEV on the study of the most massive haloes in a volume of interest, as discussed in \cite{Davis2011}. We start from the CDF given by
\begin{equation}
P_{\rm GEV}(m)\equiv {\rm Prob.}(m_{\rm max} \le m)
\equiv \int_0^{m}p_{\rm GEV}(m_{\rm max})\,\dd m_{\rm max},
\label{eq:A1}
\end{equation}
In the following we will use both, the mass $m$ and $u\equiv\log_{10}\,m$ as variables and decide case by case which is more convenient. The probability in equation~\eqref{eq:A1} to find a maximum halo mass smaller than $m$ should equal the probability $P_0(m)$ to find no halo at all with a mass bigger than $m$. Thus the probably density function (PDF) $p_{\rm GEV}(m)$ is given by
\begin{equation}
p_{\rm GEV}(m)=p_0(m)=\frac{\dd\, P_0}{\dd\, m}.
\label{eq:A3}
\end{equation}
If the volume of interest is large enough such that haloes can be considered to be unclustered, then $P_0(m)$ can be modelled by Poisson statistic \citep{Davis2011}
\begin{equation}
P_0(m)=\frac{\lambda^k \exp{\left(-\lambda\right)}}{k!}=\exp\left[-n( > m) V\right],
\label{eq:A4}
\end{equation}
where $\lambda=n( > m)V$ is the expected number of haloes more massive than $m$, thus $n( > m)$ is the comoving number density of haloes above mass $m$ and $V$ is the comoving volume. The number of occurrences $k$ is in the current case zero, leading to the last equality.\\
The parameters $\alpha, \beta$ and $\gamma$ are obtained by performing a  Taylor expansion of the two CDFs $P_0$ and $P_{\rm GEV}$ from the equations~\eqref{eq:B1} and \eqref{eq:A4} around the maxima of the distributions $\dd\,P_0/\dd\,u$ and $\dd\,P_{\mathrm{GEV}}/\dd\,u$, corresponding to the respective density functions. This gives
\begin{align*}
 P_0(u)&=P_0(u_0)+\left.\frac{\dd\,P_0(u)}{\dd\,u}\right|_{u_0} \left(u-u_0\right)+\dots\; ,\\
 P_{\mathrm{GEV}}(u)&=P_{\mathrm{GEV}}(u_0)+\left.\frac{\dd\,P_{\mathrm{GEV}}(u)}{\dd\,u}\right|_{u_0} \left(u-u_0\right)+\dots\; ,
\end{align*}
where the $u_0=\log_{10}m_0$ denotes the value of $u$ at which the density functions have their maximum. Now, we equal the first two terms of the two expansions, the second order derivatives vanish by definition and higher orders we neglect, hence one finds\footnote{When calculating $\dd\,P_0/\dd\,m$ one should keep in mind that $\dd\,n(>m)/\dd\,m=-\dd\,n/\dd\,m$.}
\begin{align}
n(>m_0)V &= \left[1+\gamma\frac{\left(u_0-\alpha\right)}{\beta}\right]^{-\frac{1}{\gamma}}, \label{eq:term1} \\
m_0 \ln{10} \left.\frac{\dd\,n}{\dd\,m}\right|_{m_0}V &= \frac{1}{\beta}\left[1+\gamma\frac{\left(u_0-\alpha\right)}{\beta}\right]^{-\frac{1}{\gamma}-1},\label{eq:term2}
\end{align}
where $\left.\dd\,n/\dd\,m\,\right|_{m_0}$ is the halo mass function evaluated at $m_0$. Together with the relation for the mode (also known as the position of the peak) given by
\begin{equation}
 u_0=\alpha+\frac{\beta}{\gamma}\left[\left(1+\gamma\right)^{-\gamma}-1\right],
\end{equation}
one arrives finally at the equations for the three parameters
\begin{eqnarray}
\gamma = n(>m_0)V-1, \quad \beta =
\frac{(1+\gamma)^{(1+\gamma)}}{\left.\frac{\dd\,n}{\dd\,m}\right|_{m_0}Vm_0\ln 10}, \nonumber \\
\alpha = \log_{10} m_0 - \frac{\beta}{\gamma}[(1+\gamma)^{-\gamma} -1]\;.
\end{eqnarray}
\begin{figure}
\centering
\includegraphics[width=0.99\linewidth]{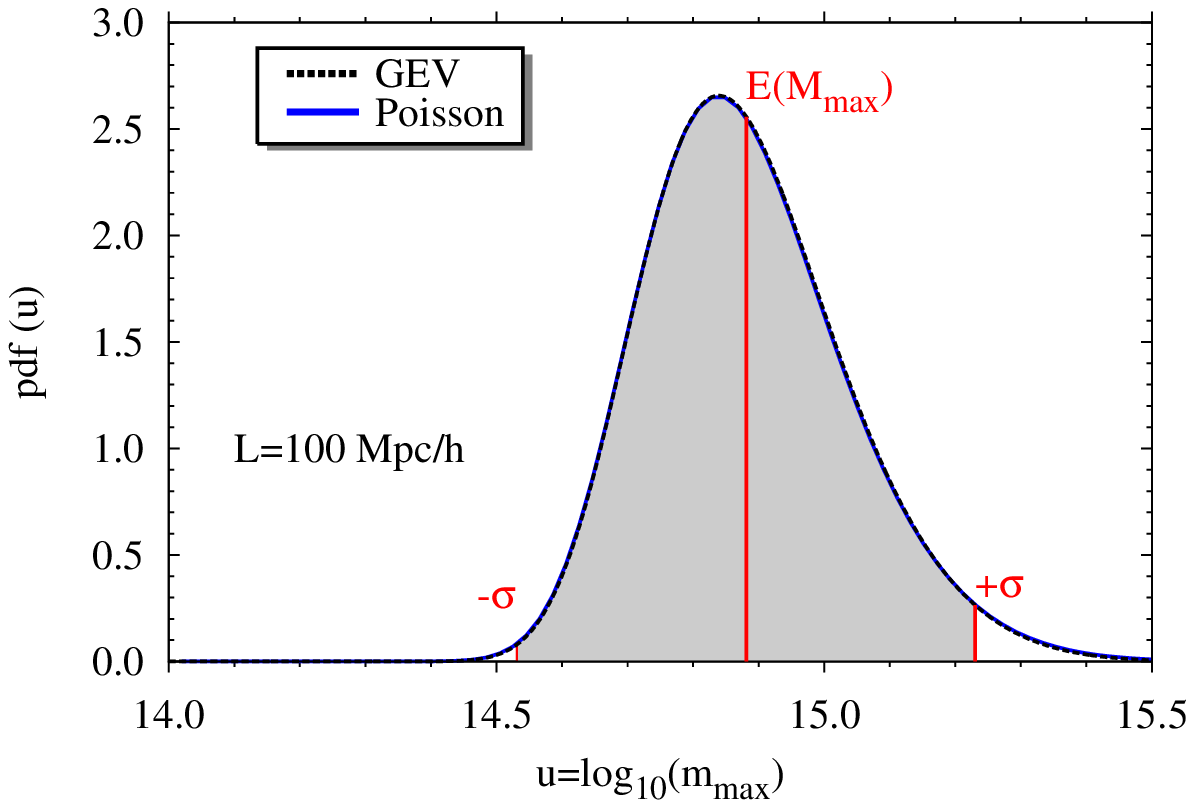}
\caption{Probability density function of the GEV and Poisson statistics for a sphere of radius $L=100\,\mathrm{Mpc}\,h^{-1}$ at $z=0$ as a function of $u=\log_{10} m_{\rm max}$. The gray shaded area denotes the $1$-$\sigma$ confidence region and the red line next to the peak (also known as mode of the distribution) is the expected value.}\label{fig:A1}
\end{figure}
The most likely mass $m_0$ is obtained via the extremal condition 
\begin{equation}
 \left.\frac{\dd^2\,P_0}{\dd\,u^2}\right|_{u_0}=\left(\ln10\right)^2 m \left(\frac{\dd\,P_0}{\dd\,m} +m\frac{\dd^2\,P_0}{\dd\,m^2}\right)=0.
\end{equation}
which can be recast into
\begin{equation}
 \frac{\dd\,n}{\dd\,m}+m\frac{\dd ^2\,n}{\dd\,m^2}+mV\left(\frac{\dd\,n}{\dd\,m}\right)^2=0,
\label{eq:cond_min0}
\end{equation}
by using the definition of $P_0$ from equation~\eqref{eq:A4}. When the mass function of \cite{Sheth&Tormen1999} (ST) is incorporated in the above relation it is possible to obtain an analytic relation for $m_0$ \citep{Davis2011}. The ST mass function is given by
\begin{equation}
 \frac{\dd\,n}{\dd\,m}=\frac{\bar{\rho}_{\rm m}}{m^2}\frac{\dd\ln\nu}{\dd\ln m}A\sqrt{\frac{a\nu}{2\pi}}\left[1+\left(a\nu\right)^{-p}\right]\e^{-\frac{a\nu}{2}},
\end{equation}
where $\bar{\rho}_{\rm m}=\Omega_{\rm m0}\,\rho_{\rm crit}$ is the mean matter density today, $\nu=\left[\delta_{\rm c}/\sigma(m,z)\right]^2$ and the ST-parameters are $A=0.3222$, $a=0.707$ and $p=0.3$. The analytic expression for $m_0$ is then found to be
\begin{equation}
\begin{split}
&A \frac{\bar{\rho}_{\rm m} V}{m_0}\sqrt{\frac{a}{2\pi \nu_0}}
e^{-a\nu_0/2} \left[ 1+(a\nu_0)^{-p} \right]  \\
& \quad { } -\frac{a}{2} - \frac{1}{2\nu_0} - \frac{ap
 (a\nu_0)^{-(p+1)}}{1+(a\nu_0)^{-p}} + \frac{\nu_0''}{\nu_0'^2} =0.
\end{split}
\end{equation}
where $\nu_0$ is $\nu$ evaluated at $m_0$ and primes denote derivatives with respect to mass.
\section{Useful relations}\label{sec:B}
In this section we summarise for the reader the most important relations for the GEV statistic as needed for this work.
\begin{itemize}
\item Cumulative distribution function (CDF)
\begin{equation}\label{eq:B1}
 P_{\mathrm{GEV}}(u) = \exp{\left\lbrace -\left[1+\gamma \left(\frac{u-\alpha}{\beta}\right)\right]^{-1/\gamma}\right\rbrace}.
\end{equation}
\item Probability density function (PDF)
\begin{equation}\label{eq:B2} 
\begin{split} 
p_{\mathrm{GEV}}(u)& = \frac{1}{\beta}\left[1+\gamma \left(\frac{u-\alpha}{\beta}\right)\right]^{-1-1/\gamma} \\
& \quad \times\exp{\left\lbrace -\left[1+\gamma \left(\frac{u-\alpha}{\beta}\right)\right]^{-1/\gamma}\right\rbrace}.
\end{split} 
\end{equation}
\item Mode - (most likely value)
\begin{equation}
 u_0=\alpha+\frac{\beta}{\gamma}\left[\left(1+\gamma\right)^{-\gamma}-1\right].
\end{equation}
\item Expected value
\begin{equation}
\mathrm{E}_{\rm GEV}=\alpha-\frac{\beta}{\gamma}+\frac{\beta}{\gamma}\Gamma\left(1-\gamma\right).
\end{equation}
\item Variance
\begin{equation}
\mathrm{VAR}_{\rm GEV}=\frac{\beta^2}{\gamma^2}\left[\Gamma\left(1-2\gamma\right)-\Gamma^2\left(1-\gamma\right)\right],
\end{equation}
 \end{itemize}
 where $\Gamma$ denotes the Gamma function. All the equations given above are valid for $\gamma<0$, which is the case for the applications in this work.
\end{document}